\documentclass[12pt,a4paper]{article}

\usepackage{geometry}
\usepackage{hyperref}
\usepackage{cite}
\usepackage{url}
\usepackage{amsmath}
\usepackage{amssymb}
\usepackage[dvips]{graphicx}
\usepackage{latexsym}
\usepackage{pstricks}
\usepackage{bm}
\usepackage{pbox}
\usepackage{placeins}
\usepackage{graphicx}% Include figure files
\usepackage{caption}
\usepackage{subcaption}
\usepackage[T1]{fontenc}
\usepackage{footnote}
\usepackage{pdfpages}
\usepackage{hhline}
\usepackage{multirow}
\usepackage{enumitem}
\usepackage[toc,page]{appendix}
\newcommand{\be}{\begin{equation}}
\newcommand{\ee}{\end{equation}}

\newcommand{\bea}[1]{\begin{eqnarray}\label{#1} }
\newcommand{\eea}{\end{eqnarray}}

\newcommand\scalemath[2]{\scalebox{#1}{\mbox{\ensuremath{\displaystyle #2}}}}

\def\beqn{\begin{eqnarray}}
\def\eeqn{\end{eqnarray}}

\def\beq{\begin{equation}}
\def\eeq{\end{equation}}
\def\bea{\begin{eqnarray}}
\def\eea{\end{eqnarray}}

\def\non{\nonumber}

\def\vs{\vspace}

\newcommand{\noi}{\noindent}
\newcommand{\bem}{\begin{pmatrix}}
\newcommand{\eem}{\end{pmatrix}}

\begin{document}
\vspace*{-0.2in}
\begin{flushright}
OSU-HEP-16-02\\
%hep-ph/
\end{flushright}

\vs{0.5cm}

\begin{center}
{\Large\bf New Class of SO(10) Models for Flavor}\\
\end{center}

\vspace{0.5cm}
\begin{center}
{\large
{}~K.S. Babu$^{a,}$\footnote{E-mail: babu@okstate.edu},{}~
Borut Bajc$^{b,}$\footnote{E-mail: borut.bajc@ijs.si} and
{}~Shaikh Saad$^{a,}$\footnote{E-mail: shaikh.saad@okstate.edu}
}
\vspace{0.5cm}

\centerline{$^{a}${\it\small Department of Physics, Oklahoma State University, Stillwater, OK, 74078, USA }}
\centerline{$^{b}${\it\small Jo\v{z}ef Stefan Institute, 1000 Ljubljana, Slovenia}}
\end{center}
%\vspace{0.6cm}

%%%%%%%%%%%%%%%%%%%%%%%%%%%%%%%%%%%%%%%%%%%%%
%%%%%%%%%%%%%%%%%%%%%%%%%%%%%%%%%%%%%%%%%%%%%
\begin{abstract}
We present a new class of unified models based on $SO(10)$ symmetry which provides insights into the masses and mixings
of quarks and leptons, including the neutrinos.  The key feature of our proposal is the absence of Higgs boson $10_H$
belonging to the fundamental representation that is normally employed.  Flavor mixing is induced via vector-like fermions in the
${16} + \overline{16}$ representation. A variety of scenarios, both supersymmetric and otherwise, are analyzed involving
a $\overline{126}_H$ along with either a $45_H$ or a $210_H$ of Higgs boson employed for symmetry breaking.  It is shown that this
framework, with only a limited number of parameters, provides an excellent fit to the full fermion spectrum, utilizing either
type-I or type-II seesaw mechanism. These flavor models can be potentially tested and distinguished in their predictions for proton
decay branching ratios, which are analyzed.
\end{abstract}

\newpage
%%%%%%%%%%%%%%%%%%%%%%%%%%%%%%%%%%%%%%%%%%%%%
%%%%%%%%%%%%%%%%%%%%%%%%%%%%%%%%%%%%%%%%%%%%%
\section{Introduction}

Grand unified theories \cite{gut1,gut2,gut3}
based on $SO(10)$ gauge symmetry \cite{so10} are attractive candidates for physics beyond the Standard Model (SM).
These theories predict the existence of right-handed neutrinos needed for the seesaw mechanism, and unify all fermions of a given
family into a single irreducible multiplet, the 16--dimensional spinor representation.  Quarks and leptons are thus
unified, as are the three gauge interactions of the SM.  The unification of fermions into multiplets
suggests that $SO(10)$ may serve as a fertile ground for understanding the flavor puzzle.  There are challenges involved, since in particular,
large neutrino mixing angles should emerge from the same underlying Yukawa structure
that allows for small quark mixing angles. This indeed has been realized in a class of $SO(10)$ models with a minimal set of Yukawa coupling
matrices \cite{Babu:1992ia,Fukuyama:2002ch,Goh:2003sy,Goh:2003hf,
Bertolini:2004eq,Babu:2005ia,Joshipura:2011nn,Altarelli:2013aqa,Dueck:2013gca}, and we shall provide a new class of models that achieves this in this paper. Since $SO(10)$ admits an intermediate symmetry, the Pati-Salam symmetry $SU(4)_c \times
SU(2)_L \times SU(2)_R$ or one of its subgroups,  unification of gauge couplings can occur consistently even without low energy supersymmetry.
Of course, $SO(10)$ may be realized in the supersymmetric context as well, in which case the intermediate symmetry breaking scale
may be the same as the unification scale.  As far as the Yukawa sector of the theory is concerned, the two scenarios (non-SUSY versus
SUSY) are not all that different.  In this paper we shall study a new class of $SO(10)$ models addressing the flavor puzzle both in the
non-supersymmetric and in the SUSY contexts.

One of our motivations for the present study is the difficulty faced by a widely studied
minimal renormalizable supersymmetric SO(10) \cite{Clark:1982ai,Aulakh:1982sw,Aulakh:2003kg}
grand unified theory.  This theory has attracted much attention in the past due to several attractive features which include:
\begin{itemize}
\item
natural generation of neutrino masses and mixings through type I \cite{Minkowski:1977sc}
and type II \cite{Magg:1980ut} seesaw mechanism;

\item
relation between neutrino and charged fermion mass matrices \cite{Babu:1992ia};

\item
good fit of fermion masses and mixings with an economic Yukawa sector with only two
symmetric Yukawa matrices \cite{Babu:1992ia,Fukuyama:2002ch,Goh:2003sy,Goh:2003hf,
Bertolini:2004eq,Babu:2005ia,Joshipura:2011nn,Altarelli:2013aqa,Dueck:2013gca};

\item
automatic and exact low energy R-parity conservation leading to a compelling dark matter candidate
\cite{mohap,Aulakh:1997ba,Aulakh:1998nn,Aulakh:1999cd,Aulakh:2000sn};

\item
connection of the $b-\tau$ Yukawa unification and large atmospheric mixing angle in
scenarios with dominant type II seesaw mechanism \cite{Bajc:2001fe,Bajc:2002iw,Bajc:2004fj}.
\end{itemize}

The Yukawa sector of this theory has only two symmetric matrices (in flavor space), involving
a $10_H$ and a $\overline{126}_H$ of Higgs bosons.  It is natural to include a $210_H$ for
completing the symmetry breaking.  In such a scenario,
unfortunately, once the constraints from the Higgs sector are properly taken into account,
the model can be ruled out \cite{Aulakh:2005bd,Bajc:2005qe,Aulakh:2005mw,Bertolini:2006pe},
assuming that the low energy supersymmetric threshold corrections to the fermion masses are negligible.
With the relatively large Higgs mass $m_H=125$ GeV,  the split
supersymmetric scenario \cite{Giudice:2004tc,ArkaniHamed:2004yi} of the minimal $SO(10)$ model
\cite{Bajc:2008dc} is also found to be inconsistent \cite{Giudice:2011cg,Bagnaschi:2014rsa}\footnote{BB thanks
Ketan Patel for pointing this out.}.

One should not abandon the whole elegant grand unified program simply because the
simplest supersymmetric realization does not work perfectly. The usual way to rule in a theory that
was ruled out is to increase the particle content and thus the number of model parameters. This was the
approach of \cite{Aulakh:2013lxa}, where a new 120-dimensional Higgs representation has been
added to the minimal model.\footnote{Another possibility, not yet fully explored, is to increase the
Higgs sector parameters, for example with a $54_H$, see Ref \cite{Goh:2003sy} for fermion fits.}
In this way the Yukawa sector increases by one antisymmetric matrix, which gives sufficient freedom
to fit the data.

In this paper we will go, surprisingly, in the opposite direction, and ask ourselves, if it is
possible to fit the data with less, not more, Yukawa matrices. This paradoxical question has
obviously a hidden proviso, otherwise we would get no mixing at all. To account for the correct
low energy mass spectrum, mixings, and CP violation we will thus make use of an extra vector-like
generation $16 + \overline{16}$, similar to the one used in \cite{Babu:1995fp}. The difference with \cite{Babu:1995fp} is that we
will assume the bilinear spinors $16_a$ to be coupled with $\overline{126}_H$ instead of $10_H$.
In this way we may hope to describe neutrino masses and mixings in a pattern similar to the
charged fermions, which is one of the great achievements of the $SO(10)$ framework.

We shall see that this decreasing of the number of Yukawa matrices at the expense of an extra
vector-like family can be successful and we will show several examples where
it works. Although we will consider different possible Higgs sectors and take some of their
constraints seriously, we will not consider a combined fit of the Higgs and Yukawa parameters,
which can obviously pose extra restrictions. This more modest approach nevertheless
shows that $SO(10)$ Yukawa sectors with a single Yukawa matrix can be realistic.

The rest of the paper is organized as follows.  In Sec. II we present the key features of the
new class of $SO(10)$ models.  In Sec. III we set up the framework and the formalism.  In Sec. IV
we adopt a specific basis that removes redundancies, which is well suited for numerical analysis
of the flavor observables.  Sec. V discusses the constraints imposed on the SUSY models from
the minimization of the Higgs potential.  Sec. VI
has our numerical fits to the fermion masses and mixings for
the six models analyzed.  Finally, in Sec. VII we conclude.  In Appendices A and B some useful relations
used for the fermion mass fits are given. Appendix C contains the numerical Yukawa matrices for various
cases that result from the fits.

\section{New class of \boldmath{$SO(10)$} models}

The key feature of the new proposed models is the absence of $10_H$.  In its place we introduce a $16+\overline{16}$
vector-like fermions.  In addition to a $\overline{126}_H$, we employ either a $45_H$ or a $210_H$ for symmetry
breaking.  These fields have non-trivial couplings to the vector-like fermions, which is needed to avoid certain
unwanted relations among down-type quark and charged lepton masses.  Additional Higgs fields (e.g. $54_H$) are
needed for consistent GUT symmetry breaking, but these fields do not enter into the Yukawa sector.  The Yukawa Lagrangian
of our models has a very simple form,
\begin{equation}
{\cal L}_{\rm Yuk} = \overline{16}\left(m_a+\eta_a45_H \right)16_a +  16_a\,{\cal Y}_{ab}\,\overline{126}_H \,16_b+\overline{16}\,\bar y\,126_H \,\overline{16} + h.c.
\label{eq1}
\end{equation}
corresponding to the use of $45_H$ as the symmetry breaking field (in addition to the $\overline{126}_H$ field).
Here $a,\,b = 1-4$ are the generation indices which include a $16$ from the vector-like family.  We thus see that the Yukawa
sector has one $4 \times 4$ matrix ${\cal Y}_{ab}$, and two four-vectors $m_a$ and $\eta_a$.  Since ${\cal Y}_{ab}$ can be
chosen to be diagonal and real, this amounts to $4+4+4$ flavor mixing parameters. The Yukawa coupling
$\overline{16}\,\bar y\,126_H$ does not have any effect on the light fermion masses and mixings.
While in the diagonal and real basis for
${\cal Y}_{ab}$ the vectors $m_a$ and $\eta_a$ are in general complex, these being related to GUT scale masses, one
complex combination disappears from low energy masses and mixings.  One should add to this set two (real) VEV ratios (one from the
two SM singlets of $45_H$ and one for the up-type and down-type Higgs doublet VEV ratio from the $\overline{126}_H$), and
an overall scale for the right-handed neutrino masses.  We thus see that the model has 14 real parameters and 7 phases
to fit 18 observed values among quark masses, quark mixings and CP violation, charged fermion masses, neutrino mass-squard differences and mixing
angles.  Thus these models are rather constrained, yet we show that excellent fits are obtained. It may be noted that
the minimal supersymmetric $SO(10)$ models with two symmetric Yukawa coupling matrices involving $10_H$ and
$\overline{126}_H$ have 12 real parameters and 7 phases that enter into the flavor sector.

The basic structure of Eq. (\ref{eq1}) can be realized in several other ways.  We study all such $SO(10)$ models in this paper.
The Higgs field $45_H$ in Eq. (\ref{eq1}) may be replaced by a $210_H$.  In this case, since the $210_H$ contains three SM singlet fields,
there are two ratios of VEVs from the $210_H$, which would increase the number of parameters by one.  These models may be realized with
or without low energy supersymmetry.  In the non-SUSY models, the VEVs of $45_H$ and $210_H$ are real, while in SUSY models they are
in general complex (thus increasing the phase parameters to 8).  In the SUSY models we find that although the $210_H$ has two associated
VEV ratios, only one of the two is independent, due to symmetry breaking constraints arising from the superpotential.  In SUSY versions,
additional fields other than $\overline{126}_H$ and  $210_H$ used in the Yukawa sector are often required, in order to avoid new chiral
supermultiplets that remain light and spoil unification of gauge couplings.  A summary of the models that fit into this classification
and studied here is given below.  All models contain a $\overline{126}_H$ (plus a $126_H$ in the case of SUSY), in addition to the
Higgs fields shown below.
\vspace*{-0.05in}
\begin{enumerate}[label=\textbf{A.}]
	\item Non-SUSY Model with $45_H+54_H$
\end{enumerate} \vspace{-20pt}
\begin{enumerate}[label=\textbf{B.}]
	\item Non-SUSY Model with $210_H+54_H$ or $210_H+16_H$
\end{enumerate} \vspace{-20pt}
\begin{enumerate}[label=\textbf{C.}]
	\item SUSY Model with $45_H+54_H+16_H+\overline{16}_H$
\end{enumerate}\vspace{-20pt}
\begin{enumerate}[label=  $\textbf{D.}$]
	\item SUSY Model with $210_H+54_H$
\end{enumerate} \vspace{-20pt}
\begin{enumerate}[label=\textbf{E.}]
	\item SUSY Model with $210_H+16_H+\overline{16}_H$
\end{enumerate} \vspace{-20pt}
\begin{enumerate}[label=\textbf{F.}]
	\item SUSY Model with $210_H+54_H+16_H+\overline{16}_H$
\end{enumerate}

The VEV of the SM singlet in $\overline{126}_H$ will be found {\it a posteriori} to be around $10^{13} - 10^{14}$ GeV in all models.
This has an effect on the choice of Higgs fields, especially in the SUSY models:
Very simple Higgs systems used for GUT symmetry breaking would lead to certain sub-multiplets
having mass of order $10^{11}$ GeV, which would spoil perturbative unification of gauge couplings in SUSY $SO(10)$.
The choice of ``other Higgs fields" shown above are in part guided by this not happening.  Furthermore, in some simplistic SUSY cases,
the Higgs doublet mass matrix becomes proportional to other color sector mass matrix.  Making a pair of Higgs doublets light would then
lead to making a pair of colored states light as well, which affects perturbative unification.  Such cases are avoided in the scenarios shown above.
In each of the models listed above, seesaw mechanism may be realized via either type-I or type-II chain.  Such sub-classes will be
denoted by a label I or II when needed.  Thus \textbf{AI} would refer to type-I seesaw in Model \textbf{A}, and likewise \textbf{AII} for type-II
seesaw in the same model.

Models \textbf{A} and \textbf{B} are nonsupersymmetric, while models \textbf{C--F} are supersymmetric.
For model \textbf{A}, in addition to $45_H$, a $54_H$ is needed to break $SO(10)$ down to the SM without going through an
intermediate $SU(5)$-symmetric limit. In Model \textbf{B} which uses a $210_H$, an additional field, either a $54_H$ or a $16_H$
is needed for the following reason.  As noted already, $126_H$ acquires a VEV of order $10^{13}-10^{14}$ GeV, which can be ignored
for the study of GUT symmetry breaking at around $10^{16}$ GeV.  A single $210_H$ would break $SO(10)$ down to one of its maximal
little groups, such as $SU(5) \times U(1)$, $SU(4)_C \times SU(2)_L \times SU(2)_R$ etc.  The fermion mass matrix would then
reflect this unbroken symmetry, which is not realistic in the light fermion spectrum.  Addition of a $54_H$ (or a $16_H)$ with a GUT VEV
would reduce the surviving symmetry and help with realistic fermion masses.
For SUSY $SO(10)$, it is not a viable model if the symmetry is only broken by $45_H + 54_H$, since in this case, the Higgs doublet (1,2,1/2) and the Higgs octet (8,2,1/2)  mass matrices become  identical. So one cannot make the MSSM doublet fields light without also making the octet fields light. To break this degeneracy one needs to extend the Higgs sector. For this purpose in model \textbf{C}, we enlarge the Higgs sector by adding $16_H+\overline{16}_H$.  SUSY $SO(10)$ model with $210_H+126_{H}+\overline{126}_{H}$ is also not a consistent model, because with the requirement $v_R \sim 10^{13-14}$ GeV, the octet $(8,3,0)$ Higgs field becomes light with a mass of order $\sim 10^{10-11}$ GeV, so the theory does not remain perturbative up to the GUT scale. Thus, in order to avoid this, in model \textbf{D}, we include $54_H$ Higgs and in model \textbf{E}, we
include a $16_H+\overline{16}_H$. It will be shown later in Sec. V that, in all these SUSY $SO(10)$ models with a $210_H$, there is only one independent VEV ratio involving the $210_H$ field, owing to symmetry breaking constraints.
Including more Higgs multiplets, one can break such relationships among VEVs which can lead to two independent VEV ratios for $210_H$. We also consider this general case which is labeled as model \textbf{F}, where in addition to $210_H$, one has both $54_H$ and  $16_H+\overline{16}_H$
(or some unspecified) multiplets. It is to be mentioned that, we do not consider any model where both the $45_H$ and $210_H$ are present simultaneously, which would lead to more parameters and thus less predictions in the fermion sector. Details of the symmetry breaking schemes will be explained further in Sec. V.

%%%%%%%%%%%%%%%%%%%%%%%%%%%%%%%%%%%%%%%%%%%%%%%%%%%%%%%%%%%%%%%%
%%%%%%%%%%%%%%%%%%%%%%%%%%%%
\section{The set-up and formalism}
%%%%%%%%%%%%%%%%%%%%%%%%%%%%

All models we study have one vector like $16+\overline{16}$ pair plus $3$ generations of
chiral 16's. Their mass terms and couplings to a $45_H$ given in Eq. (\ref{eq1}) can be expanded to yield
\vspace{-10pt}
\bea
\label{16bar4516}
\overline{16}\left(m_a+\eta_a45_H \right)16_a&=&
\bar L\left(m_a+\eta_a(3v_1)\right)L_a+
\bar Q\left(m_a+\eta_a(-v_1)\right)Q_a\nonumber\\
&+&e^c_a\left(m_a+\eta_a(-3v_1-v_2)\right)\bar e^c+
\nu^c_a\left(m_a+\eta_a(-3v_1+v_2)\right)\bar\nu^c\nonumber\\
&+&d^c_a\left(m_a+\eta_a(v_1-v_2)\right)\bar d^c+
u^c_a\left(m_a+\eta_a(v_1+v_2)\right)\bar u^c ,
\eea

\noi
where $a=1,\ldots,4$ and
\vspace{-10pt}
\bea
v_1=\langle 45_H \rangle_{(1,1,15)}&,&v_2=\langle 45_H \rangle_{(1,3,1)} .
\eea
These are the SM singlet components of $45_H$ which acquire GUT scale VEVs denoted here as $v_{1,2}$.

The mass terms are of the general form
\vspace{-5pt}
\beq
\bar\psi M_a\psi_a .
\eeq
Although by redefining the phases of $\psi_a$ we can make all these $M_a$ real, we will
keep them complex in general. Then we project to the heavy states as usual by
\vspace{-8pt}
\beq
\label{psi}
\psi_a\to U_{ab}\psi_b ,
\eeq

\noi
with
\vspace{-10pt}
\bea
U&=&
\bem
\Lambda &\Lambda x^* \\ -x^T\Lambda & \bar\Lambda
\eem\hskip 1cm (U^\dagger=U^{-1})\\
\Lambda&=&1-\frac{x^*x^T}{\sqrt{1+|x|^2}(\sqrt{1+|x|^2}+1)}\hskip 1cm (\Lambda^\dagger=\Lambda)\\
x^T&=&\frac{1}{M_4}(M_1,M_2,M_3)\hskip 0.5cm,\hskip 0.5cm
\bar\Lambda=\frac{1}{\sqrt{1+|x|^2}} .
\eea

To this we add the Yukawa couplings to $\overline{126}_H$. Although we are
free to choose this $4 \times 4$ Yukawa matrix to be diagonal and real (in the original basis, i.e. before (\ref{psi})),
we will keep it to be complex symmetric and choose a convenient basis later on.
The $\overline{16}$ has coupling  to the $126_H$, but this will turn out to not affect light fermion masses.
The relevant Yukawa couplings are (see Eq. (\ref{eq1}))
\vspace{-5pt}
\beq
\label{yukawa}
16_a\,{\cal Y}_{ab}\,\overline{126}_H \,16_b+\overline{16}\,\bar y\,126_H \,\overline{16} .
\eeq

In this original basis we put all together:
\vspace{-5pt}
\bea
\label{mass1}
&&\bem
d^c_a & \overline{d}
\eem
\bem
{\cal Y}_{ab}v_d  & m_a+\eta_a(v_1-v_2) \\
m_b+\eta_b(-v_1) & \bar yv_d
\eem
\bem
d_b \\ \bar d^c
\eem\nonumber\\
&+&\bem
u^c_a & \overline{u}
\eem
\bem
{\cal Y}_{ab}v_u & m_a+\eta_a(v_1+v_2) \\
m_b+\eta_b(-v_1) & \bar yv_u
\eem
\bem
u_b \\ \bar u^c
\eem\nonumber\\
&+&\bem
e^c_a & \overline{e}
\eem
\bem
-3{\cal Y}_{ab}v_d & m_a+\eta_a(-3v_1-v_2) \\
m_b+\eta_b(3v_1) & -3\bar yv_d
\eem
\bem
e_b \\ \bar e^c
\eem\\
&+&\bem
\nu^c_a & \overline{\nu}
\eem
\bem
-3{\cal Y}_{ab}v_u & m_a+\eta_a(-3v_1+v_2) \\
m_b+\eta_b(3v_1) & -3\bar yv_u
\eem
\bem
\nu_b \\ \bar\nu^c
\eem\nonumber\\
&+&\frac{1}{2}
\bem
\nu^c_a & \bar\nu
\eem
\bem
{\cal Y}_{ab}v_R\ & 0 \\
0 & \bar y\bar v_L
\eem
\bem
\nu^c_b \\ \bar\nu
\eem+\frac{1}{2}
\bem
\nu_a & \bar\nu^c
\eem
\bem
{\cal Y}_{ab}v_L & 0 \\
0 & \bar y\bar v_R
\eem
\bem
\nu_b \\ \bar\nu^c
\eem ,
\nonumber
\eea

\noi
where
\vspace{-12pt}
\bea
v_R=\langle\overline{126}_H \rangle_{(1,3,10)}&,&
v_L=\langle\overline{126}_H \rangle_{(3,1,\overline{10})}\nonumber\\
\bar v_R=\langle 126_H \rangle_{(1,3,\overline{10})}&,&
\bar v_L=\langle 126_H \rangle_{(3,1,10)}\\
v_u=\langle\overline{126}_H \rangle_{(2,2,15)_u}&,&
v_d=\langle\overline{126}_H \rangle_{(2,2,15)_d} .\nonumber
\eea
Here $v_R$ and $\bar v_R$ are close to, but somewhat below the GUT scale, while $v_{u,d}$ are the VEVs
of the electroweak Higgs doublets arising from the $\overline{126}_H$.  $v_L$ and $\bar v_L$ denote the induced VEVs of the
$SU(2)_L$ triplets from $126_H$ and $\overline{126}_H$. In non-supersymmetric models, we have $\bar v_R = v_R^*$, $v_d = v_u^*$ and $\bar v_L = v_L^*$.

After the transformation given in Eq. (\ref{psi}) the matrices Eq. (\ref{mass1}) become
\vspace{-3pt}
\bea
\label{mass2}
&\to&\bem
d^c_a & \overline{d}
\eem
\bem
(U^T_{d^c})_{ae}{\cal Y}_{ef}v_d (U_Q)_{fb} & M_{d^c}\delta_{a4} \\
M_Q\delta_{b4} & \bar yv_d
\eem
\bem
d_b \\ \bar d^c
\eem\nonumber\\
&+&\bem
u^c_a & \overline{u}
\eem
\bem
(U_{u^c}^T)_{ae}{\cal Y}_{ef}v_u (U_Q)_{fb} & M_{u^c}\delta_{a4} \\
M_{Q}\delta_{b4} & \bar yv_u
\eem
\bem
u_b \\ \bar u^c
\eem\nonumber\\
&+&\bem
e^c_a & \overline{e}
\eem
\bem
(U_{e^c}^T)_{ae}(-3){\cal Y}_{ef}v_d (U_L)_{fb} & M_{e^c}\delta_{a4} \\
M_{L}\delta_{b4} & -3\bar yv_d
\eem
\bem
e_b \\ \bar e^c
\eem\\
&+&\bem
\nu^c_a & \overline{\nu}
\eem
\bem
(U_{\nu^c}^T)_{ae}(-3){\cal Y}_{ef}v_u (U_L)_{fb} & M_{\nu^c}\delta_{a4} \\
M_{L}\delta_{b4} & -3\bar yv_u
\eem
\bem
\nu_b \\ \bar\nu^c
\eem\nonumber\\
&+&\frac{1}{2}
\bem
\nu^c_a & \bar\nu
\eem
\bem
(U_{\nu^c}^T)_{ae}{\cal Y}_{ef}v_R(U_{\nu^c})_{fb} & 0 \\
0 & \bar y\bar v_L
\eem
\bem
\nu^c_b \\ \bar\nu
\eem\nonumber\\
&+&\frac{1}{2}
\bem
\nu_a & \bar\nu^c
\eem
\bem
(U_L^T)_{ae}{\cal Y}_{ef}v_L(U_L)_{fb} & 0 \\
0 & \bar y\bar v_R
\eem
\bem
\nu_b \\ \bar\nu^c
\eem .
\nonumber
\eea

To get the light fermion mass matrices defined as
\vspace{-5pt}
\beq
{\cal L}=d^{cT}M_Dd+u^{cT}M_Uu+e^{cT}M_Ee+\frac{1}{2}\nu^T M_N\nu+h.c.
\eeq

\noi
we have to project to the light generations. In doing so we need to evaluate
(${\cal Y}$ is a $4\times 4$ matrix, while $Y$ is its $3\times3$ submatrix)
\vspace{-3pt}
\bea
\label{U1TYU2}
U_1^T{\cal Y}U_2&=&
\bem
\Lambda_1^T & -\Lambda_1^Tx_1 \\ x_1^\dagger\Lambda_1^T & \bar\Lambda_1
\eem
\bem
Y & y \\ y^T & y_4
\eem
\bem
\Lambda_2 & \Lambda_2x_2^* \\ -x_2^T\Lambda_2 & \bar\Lambda_2
\eem\\
&=&
\bem
\Lambda_1^T(Y-yx_2^T-x_1y^T+y_4x_1x_2^T)\Lambda_2 &
\Lambda_1^T(Yx_2^*+y-x_1y^Tx_2^*-y_4x_1)\bar\Lambda_2 \\
\bar\Lambda_1(x_1^\dagger Y+y^T-x_1^\dagger yx_2^T-y_4x_2^T)\Lambda_2 &
\bar\Lambda_1(x_1^\dagger Yx_2^*+y^Tx_2^*+x_1^\dagger Y+y_4)\bar\Lambda_2
\eem , \nonumber
\eea

\noi
where we used $\Lambda x^*=\bar\Lambda x^*$.

For charged fermions this is enough, and we get
(mass matrices are defined as $\psi^cM_\Psi\psi$)
\vspace{-5pt}
\bea
\label{MD}
M_D&=&v_d\Lambda_{d^c}^T\left(Y-yx_Q^T-x_{d^c}y^T+y_4x_{d^c}x_Q^T\right)\Lambda_Q\\
\label{MU}
M_U&=&v_u\Lambda_{u^c}^T\left(Y-yx_Q^T-x_{u^c}y^T+y_4x_{u^c}x_Q^T\right)\Lambda_Q\\
\label{ME}
M_E&=&-3v_d\Lambda_{e^c}^T\left(Y-yx_L^T-x_{e^c}y^T+y_4x_{e^c}x_L^T\right)\Lambda_L .
\eea
\noi
For neutrinos things are slightly more involved, since there are two kinds of heavy neutrinos,
the usual right-handed ones, plus the new vector-like ones. The full symmetric Majorana mass
matrix is $10\times 10$. However, in the leading order in $yv_R/M_{L,\nu^c}$ ($M_{L,\nu^c}$ denote the
masses of vector-like leptons), the situation
returns to ordinary with
\vspace{-8pt}
\bea
\label{mnd}
M_{\nu_D}&=&-3v_u\Lambda_{\nu^c}^T(Y-yx_L^T-x_{\nu^c}y^T+y_4x_{\nu^c}x_L^T)\Lambda_L\\
\label{mnr}
M_{\nu_R}&=&v_R\Lambda_{\nu^c}^T(Y-yx_{\nu^c}^T-x_{\nu^c}y^T+y_4x_{\nu^c}x_{\nu^c}^T)\Lambda_{\nu^c}\\
\label{mnl}
M_{\nu_L}&=&v_L\Lambda_{L}^T(Y-yx_L^T-x_Ly^T+y_4x_{L}x_{L}^T)\Lambda_{L} ,
\eea

\noi
so that as usual by using the seesaw ~\cite{Minkowski:1977sc} formula we arrive at the $3 \times 3$ light neutrino mass matrix as
\vspace{-3pt}
\beq
\label{mn}
M_N=M_{\nu_L}-M_{\nu_D}^TM_{\nu_R}^{-1}M_{\nu_D} .
\eeq

If the approximation $yv_R/M_{L,\nu^c} \ll 1$ is not good, we write the full symmetric matrix
 for
$(\nu_i,\nu^c_i,\nu_4,\bar\nu^c,\nu^c_4,\bar\nu)$:
\vspace{-8pt}
\beq
\bem
(U_L^T(v_L{\cal Y})U_L)_{ij} & (U_L^T(-3v_u{\cal Y})U_{\nu^c})_{ij} &
(U_L^T(v_L{\cal Y})U_L)_{i4} & 0 & (U_L^T(-3v_u{\cal Y})U_{\nu^c})_{i4} & 0 \\
(U_{\nu^c}^T(-3v_u{\cal Y})U_L)_{ij} & (U_{\nu^c}^T(v_R{\cal Y})U_{\nu^c})_{ij} &
(U_{\nu^c}^T(-3v_u{\cal Y})U_L)_{i4} & 0 & (U_{\nu^c}^T(v_R{\cal Y})U_{\nu^c})_{i4} & 0 \\
(U_L^T(v_L{\cal Y})U_L)_{4j} & (U_L^T(-3v_u{\cal Y})U_{\nu^c})_{4j} &
(U_L^T(v_L{\cal Y})U_L)_{44} & 0 & (U_L^T(-3v_u{\cal Y})U_{\nu^c})_{44} & M_L \\
0 & 0 & 0 & \bar v_Ry & M_{\nu^c} & -3v_uy \\
(U_{\nu^c}^T(-3v_u{\cal Y})U_L)_{4j} & (U_{\nu^c}^T(v_R{\cal Y})U_{\nu^c})_{4j} &
(U_{\nu^c}^T(-3v_u{\cal Y})U_L)_{44} & M_{\nu^c} & (U_{\nu^c}^T(v_R{\cal Y})U_{\nu^c})_{44} & 0 \\
0 & 0 & M_L & -3v_uy & 0 & \bar v_Ly
\eem .
\eeq

\noindent One can integrate out $\nu_4$ and $\bar\nu$ without any trace, since they mix through a
large $M_L$, but otherwise feel just the small VEVs. What remains is for
$(\nu_i,\nu^c_i,\bar\nu^c,\nu^c_4)$:

\beq
\bem
(U_L^T(v_L{\cal Y})U_L)_{ij} & (U_L^T(-3v_u{\cal Y})U_{\nu^c})_{ij} &
0 & (U_L^T(-3v_u{\cal Y})U_{\nu^c})_{i4}  \\
(U_{\nu^c}^T(-3v_u{\cal Y})U_L)_{ij} & (U_{\nu^c}^T(v_R{\cal Y})U_{\nu^c})_{ij} &
 0 & (U_{\nu^c}^T(v_R{\cal Y})U_{\nu^c})_{i4} \\
0 & 0 &  \bar v_Ry & M_{\nu^c}  \\
(U_{\nu^c}^T(-3v_u{\cal Y})U_L)_{4j} & (U_{\nu^c}^T(v_R{\cal Y})U_{\nu^c})_{4j} &
M_{\nu^c} & (U_{\nu^c}^T(v_R{\cal Y})U_{\nu^c})_{44}  \eem .
\eeq

\noindent This has again the form
\vspace{-3pt}
\beq
\bem
M_{\nu_L} & M_{\nu_D}^T \\ M_{\nu_D} & M_{\nu_R}
\eem .
\eeq

\noi
and thus Eq. (\ref{mn}) applies with $M_{\nu_L}$ given by Eq. (\ref{mnl}), but now for
$5$ right-handed neutrinos with a $5\times 3$ matrix $M_{\nu_D}$ and a
$5\times 5$ symmetric matrix $M_{\nu_R}$:
\vspace{-5pt}
\begin{align}
M_{\nu_D}&=
(-3v_u)\bem
\Lambda_{\nu^c}(Y-yx_L^T-x_{\nu^c}y^T+y_4x_{\nu^c}x_L^T)\Lambda_L \\
0 \\
\bar\Lambda_{\nu^c}(x_{\nu^c}^TY+y^T-x_{\nu^c}^Tyx_L^T-x_L^Ty_4)\Lambda_L
\eem\\
M_{\nu_R}&=
\bem
v_R\Lambda_{\nu^c}(Y-yx_{\nu^c}^T-x_{\nu^c}y^T+y_4x_{\nu^c}x_{\nu^c}^T)\Lambda_{\nu^c}&
 0 &
v_R\Lambda_{\nu^c}(Yx_{\nu^c}+y-x_{\nu^c}y^Tx_{\nu^c}-y_4x_{\nu^c}) \bar\Lambda_{\nu^c}\\
0 &  \bar v_Ry & M_{\nu^c}  \\
v_R\bar\Lambda_{\nu^c}(x_{\nu^c}^TY+y^T-x_{\nu^c}^Tyx_{\nu^c}^T-y_4x_{\nu^c}^T)\Lambda_{\nu^c}&
M_{\nu^c} &
v_R\bar\Lambda_{\nu^c}(x_{\nu^c}^TYx_{\nu^c}+y^Tx_{\nu^c}+x_{\nu^c}^Ty+y_4)\bar\Lambda_{\nu^c}
\eem .
\end{align}

To conclude, let's write down explicitly the various $\vec{x}$'s:
\vspace{-3pt}
\bea
\vec{x}_L=\frac{\vec{m}+\vec{\eta}(3v_1)}{m_4+\eta_4(3v_1)}&,&
\vec{x}_Q=\frac{\vec{m}+\vec{\eta}(-v_1)}{m_4+\eta_4(-v_1)} ,\\
\vec{x}_{e^c}=\frac{\vec{m}+\vec{\eta}(-3v_1-v_2)}{m_4+\eta_4(-3v_1-v_2)}&,&
\vec{x}_{\nu^c}=\frac{\vec{m}+\vec{\eta}(-3v_1+v_2)}{m_4+\eta_4(-3v_1+v_2)} ,\\
\vec{x}_{d^c}=\frac{\vec{m}+\vec{\eta}(v_1-v_2)}{m_4+\eta_4(v_1-v_2)}&,&
\vec{x}_{u^c}=\frac{\vec{m}+\vec{\eta}(v_1+v_2)}{m_4+\eta_4(v_1+v_2)} .
\eea

Defining
\vspace{-3pt}
\beq
\label{xuy}
\vec{x}=\frac{\vec{m}}{m_4}\hskip 0.5cm,\hskip 0.5cm
u_{1,2}=\eta_4\frac{v_{1,2}}{m_4}\hskip 0.5cm,\hskip 0.5cm
\vec{z}=\frac{\vec{\eta}}{\eta_4} ,
\eeq

\noi
we can rewrite the above as
\vspace{-3pt}
\bea
\vec{x}_L=\frac{\vec{x}+\vec{z}(3u_1)}{1+(3u_1)}&,&
\vec{x}_Q=\frac{\vec{x}+\vec{z}(-u_1)}{1+(-u_1)} ,\non\\
\label{x}
\vec{x}_{e^c}=\frac{\vec{x}+\vec{z}(-3u_1-u_2)}{1+(-3u_1-u_2)}&,&
\vec{x}_{\nu^c}=\frac{\vec{x}+\vec{z}(-3u_1+u_2)}{1+(-3u_1+u_2)} ,\\
\vec{x}_{d^c}=\frac{\vec{x}+\vec{z}(u_1-u_2)}{1+(u_1-u_2)}&,&
\vec{x}_{u^c}=\frac{\vec{x}+\vec{z}(u_1+u_2)}{1+(u_1+u_2)} .\non
\eea

To get the masses and mixings we change the basis
\vspace{-5pt}
\beq
\label{XLXR}
x\to X_Lx\;\;\;,\;\;\;x^c\to X_R^*x^c
\eeq

\noi
for $x=d,u,e,\nu$ and $X=D,U,E,N$. This means that (for $X=N$, $X_R=X_L^{\ast}$)
\vspace{-5pt}
\beq
M_X=X_RM_X^dX_L^\dagger
\eeq

\noi
so that the CKM and PMNS matrices are defined as
\vspace{-8pt}
\bea
V_{CKM}&=&U_L^\dagger D_L\\
V_{PMNS}&=&E_L^\dagger N_L .
\eea

So far we have been very general.  However, there are redundancies that are present, which
should be removed for an efficient numerical fitting algorithm.
In the next section we shall choose a specific basis, which may appear at first to be
less intuitive but which is well-suited for our numerical minimization.  There are two
obvious basis choices, one where ${\cal Y}_{ab}$ is diagonal, and a second one where
the vectors $m_a$ and $\eta_a$ have simple forms.  It is the second one that is used in
the next section.   For further use we give here the relations between the two sets of
parameters.
\vspace{-5pt}
\bea
 \label{eq:45O1}
\vec{x}&=&(0,0,\tan{\theta}e^{-i\phi})  \\
 \label{eq:45O2}
\vec{z}&=&(0,0,0)  \\
Y_{ij}&=&a_{ij} \label{eq:45O3}  \\
\label{eq:45O4}
y&=&(a_{41},a_{42},a_{43})   \\
 \label{eq:45O5}
y_4&=&a_{44}
\eea

\noi
and
\vspace{-5pt}
\bea
\label{u1-45}
u_1&=&-\frac{Te^{-i\phi}}{\cos{\theta}}\frac{\epsilon}{5}\\
\label{u2-45}
u_2&=&-\frac{Te^{-i\phi}}{\cos{\theta}}\left(1+\frac{3\epsilon}{5}\right) .
\eea

%%%%%%%%%%%%%%%%%%%%%%%%%%%%%%%%%%%%%%%
%%%%%%%%%%%%%%%%%%%%%%%%%%%%%%%%%%%%%%%
\subsection{$210_H$ instead of $45_H$}\label{section:21045}

If the $45_H$ is replaced by a $210_H$, we simply change Eq. (\ref{16bar4516}) into:
\vspace{-5pt}
\bea
\label{16bar210_H 16}
\overline{16}\left(m_a+\eta_a210\right)16_a&=&
\bar L\left(m_a+\eta_a(\phi_1-3\phi_2)\right)L_a\non\\
&+&\bar Q\left(m_a+\eta_a(\phi_1+\phi_2)\right)Q_a\non\\
&+&e^c_a\left(m_a+\eta_a(-\phi_1-3\phi_2+3\phi_3)\right)\bar e^c\\
&+&\nu^c_a\left(m_a+\eta_a(-\phi_1-3\phi_2-3\phi_3)\right)\bar\nu^c\non\\
&+&d^c_a\left(m_a+\eta_a(-\phi_1+\phi_2-\phi_3)\right)\bar d^c\non\\
&+&u^c_a\left(m_a+\eta_a(-\phi_1+\phi_2+\phi_3)\right)\bar u^c\non
\eea

\noindent where
\vspace{-5pt}
\beq \label{eq:210vev}
\phi_1=\langle210_H \rangle_{(1,1,1)}\;\;\;,\;\;\;
\phi_2=\langle210_H \rangle_{(1,1,15)}\;\;\;,\;\;\;
\phi_3=\langle210_H \rangle_{(1,3,15)}
\eeq
are the VEVs of the three SM singlets of $210_H$.

This then changes Eq. (\ref{x}) into
\bea
\vec{x}_L=\frac{\vec{x}+\vec{z}(u_1-3u_2)}{1+(u_1-3u_2)}&,&
\vec{x}_Q=\frac{\vec{x}+\vec{z}(u_1+u_2)}{1+(u_1+u_2)} ,\non\\
\vec{x}_{e^c}=\frac{\vec{x}+\vec{z}(-u_1-3u_2+3u_3)}{1+(-u_1-3u_2+3u_3)}&,&
\vec{x}_{\nu^c}=\frac{\vec{x}+\vec{z}(-u_1-3u_2-3u_3)}{1+(-u_1-3u_2-3u_3)} ,\\
\vec{x}_{d^c}=\frac{\vec{x}+\vec{z}(-u_1+u_2-u_3)}{1+(-u_1+u_2-u_3)}&,&
\vec{x}_{u^c}=\frac{\vec{x}+\vec{z}(-u_1+u_2+u_3)}{1+(-u_1+u_2+u_3)} , \non
\eea

\noi
where now
\vspace{-5pt}
\beq
u_{1,2,3}=\eta_4\frac{\phi_{1,2,3}}{m_4} .
\eeq

For correspondence with the specific basis chosen in the next section,
we still have Eqs. (\ref{eq:45O1})- (\ref{eq:45O5}), but
Eqs. (\ref{u1-45})-(\ref{u2-45}) are replaced by
\vspace{-5pt}
\bea \label{eq:210O1}
u_1&=&\frac{Te^{-i\phi}}{\cos{\theta}}\\
u_2&=&\frac{Te^{-i\phi}}{\cos{\theta}}\frac{\epsilon_1}{\sqrt{3}}\\
u_3&=&\frac{Te^{-i\phi}}{\cos{\theta}}\epsilon_2\sqrt{\frac{2}{3}}~.
\eea

%%%%%%%%%%%%%%%%%%%%%%%%%%%%%%%%%%%%%%%
%%%%%%%%%%%%%%%%%%%%%%%%%%%%%%%%%%%%%%%
\section{Analysis in a specific basis}\label{section:charged}

The general formulas given in the previous section for the light fermion mass matrices have built-in redundancies. Here we choose a specific
basis where these redundancies are removed.  We choose a basis where the four-vectors in Eq. (\ref{eq1}) have simple forms:
\begin{eqnarray}
\eta_{a}= (0,0,0,1)\,b,~~m_{a}= (0,0,\sin\theta,e^{i \phi} \cos\theta)\, M~.
\end{eqnarray}
These simple forms are achieved by $4 \times 4$ family rotation, which makes the vector $\vec{\eta}$ to have the form shown, and a subsequent
$3 \times 3$ family rotation that brings the vector $\vec{m}$ to this form.  A further rotation in the first two family space can be made, we
choose this rotation to make the $4 \times 4$ Yukawa matrix, denoted as $a_{ij}$ in this specific basis, to be diagonal in the 1-2 subspace,
i.e., $a_{12} = a_{21} = 0$.  The correspondence given in Eqs. (\ref{eq:45O1})- (\ref{eq:45O5}) as well as Eqs. (\ref{u1-45})-(\ref{u2-45})
for the case of $45_H$ arise from this choice of basis.  (The parameters $T$ and $\epsilon$ will be defined shortly.)  Let us denote
$\Phi = 45_H$ or $210_H$ and the VEV of $\Phi$ to be $\left \langle \Phi \right \rangle = \Omega$ which has two components (for $\Phi = 45_H$)
or three components (for $\Phi = 210_H$).  The Yukawa Lagrangian in this specific basis takes the form:
\vspace{-10pt}
\begin{equation}
\mathcal{L}_{\rm Yuk} = \sum^{4}_{i,j=1} a_{ij}\; 16_{i}\; 16_{j}\; \overline{126}_{H} +  \bar{y}\; \overline{16}\; \overline{16}\; 126_{H} +    b\; \overline{16}\; 16_{4}\; \Phi +   M\; \overline{16}\; (\sin\theta \; 16_{3} + e^{i \phi} \cos\theta \; 16_{4} ) .
\label{eq2}
\end{equation}
The effective mass terms that arise after the VEV of $\Phi$ is inserted would depend on the VEV ratio of the two SM singlets in $45_H$
and on two VEV ratios of the three SM singlets in the case of $210_H$.  For the former, we can define an unbroken charge $Q$, which is
not the electric charge, but a linear combination of hypercharge $Y$ and the $U(1)_X$ charge contained in $SO(10) \rightarrow SU(5) \times
U(1)_X$  -- the $45_H$ leaves this charge $Q$ unbroken.  A parameter $\epsilon$ can be introduced in terms of which
the unbroken charge $Q$ can be defined for each of the SM fermions  \cite{Babu:1995fp}:

\vspace{-8pt}
\begin{equation}
Q=-\frac{1}{5} X + \frac{6 (\epsilon+1)}{5} \frac{Y}{2} = 2 I_{3R} + \frac{6 \epsilon}{5} \frac{Y}{2} ,
\end{equation}

\noindent
where $X$ is normalized so that $X_{10 \in 16}=1$, $X_{\overline{5} \in 16}=-3$ and $X_{1 \in 16}=5$. Thus the charges of fermions $\in 16$ of $SO(10)$ for the case of $45_{H}$ are:

\vspace{-17pt}
\begin{eqnarray}\label{eq:Q45}
\begin{aligned}
Q_{u,d} = \frac{1}{5} \epsilon ; \;
Q_{u^{c}} = -1 - \frac{4}{5} \epsilon ; \;
Q_{d^{c}} = 1 + \frac{2}{5} \epsilon ;  \\
Q_{e,\nu} = -\frac{3}{5} \epsilon ; \;
Q_{e^{c}} = 1+ \frac{6}{5} \epsilon ; \;
Q_{\nu^{c}} = -1.
\end{aligned}
\end{eqnarray}

\noindent For $210_{H}$ case the fermion charges are given in terms of two parameters $\epsilon_{1,2}$:

\vspace{-25pt}
\begin{eqnarray}\label{eq:Q210}
\begin{aligned}
Q_{u,d} = 1+\frac{\epsilon_{1}}{\sqrt{3}}  ; \;
Q_{u^{c}} = -1 +\frac{\epsilon_{1}}{\sqrt{3}} + \sqrt{\frac{2}{3}} \epsilon_{2} ; \;
Q_{d^{c}} = -1 +\frac{\epsilon_{1}}{\sqrt{3}} - \sqrt{\frac{2}{3}} \epsilon_{2} ;   \\
Q_{e,\nu} = 1 -\sqrt{3} \epsilon_{1}  ; \;
Q_{e^{c}} =  -1 -\sqrt{3} \epsilon_{1} + \sqrt{6} \epsilon_{2}; \;
Q_{\nu^{c}} = -1 -\sqrt{3} \epsilon_{1} - \sqrt{6} \epsilon_{2}.
\end{aligned}
\end{eqnarray}
These charges are obtained from Eq. (\ref{16bar210_H 16}) by setting $\phi_1 = 1,\,\phi_2 = \epsilon_1/\sqrt{3}$ and $\phi_3 = \sqrt{2/3}\, \epsilon_2$.

For non-SUSY case, $Q_{f}=Q^{\ast}_{f}$ as $\Phi$ is a real field in this case, while $Q_f$ is complex in the case of SUSY. Now, writing $b \,\overline{16}\, 16_{4}\, \langle \Phi \rangle = b \Omega (  \overline{f} Q_{f}  f_{4} + \overline{f}^{c}  Q_{f^{c}} f^{c}_{4} )$, the last two terms of the Yukawa Lagrangian in Eq. (\ref{eq2}) can be written as

\vspace{-20pt}
\begin{equation}
\mathcal{L}_{\rm Yuk} \supseteq \overline{f} [ (M \sin\theta ) f_{3} + (M e^{i \phi} \cos\theta + b \Omega Q_{f}) f_{4} ]   +   \overline{f}^{c} [ (M \sin\theta ) f^{c}_{3} + (M e^{i \phi} \cos\theta + b \Omega Q_{f^{c}}) f^{c}_{4} ] .
\end{equation}

\noindent Then defining

\vspace{-12pt}
\begin{equation}
T \equiv b \Omega /M; \; N_{f,f^{c}} \equiv \sqrt{1+T^{2} |Q_{f,f^{c}}|^{2}+T \cos\theta (e^{-i \phi} Q_{f,f^{c}}+e^{i \phi} Q^{\ast}_{f,f^{c}})} ,
\end{equation}

\noindent
the heavy (GUT scale) fields ($\hat{f}_{4}, \hat{f}^{c}_{4}$) and the light SM fields ($\hat{f}_{3}, \hat{f}^{c}_{3}$) can be identified as

\vspace{-12pt}
\begin{eqnarray}
\frac{ ( \sin\theta ) f_{3} + ( e^{i \phi} \cos\theta + T Q_{f}) f_{4} }{N_{f}} &\equiv \hat{f}_{4} \;\; ;\;  \frac{  ( e^{-i \phi} \cos\theta + T Q^{\ast}_{f})  f_{3} - ( \sin\theta ) f_{4} }{N_{f}} &\equiv \hat{f}_{3} ;\\
\frac{ ( \sin\theta ) f^{c}_{3} + ( e^{i \phi} \cos\theta + T Q_{f^{c}}) f^{c}_{4} }{N_{f^{c}}}    &\equiv \hat{f}^{c}_{4} \;\; ;\; \frac{ ( e^{-i \phi} \cos\theta + T Q^{\ast}_{f^{c}}) f^{c}_{3} - ( \sin\theta )  f^{c}_{4} }{N_{f^{c}}} &\equiv \hat{f}^{c}_{3}.
\end{eqnarray}
These expressions are valid for $f=u,d,e,\nu$ and $f^c = u^c, d^c, e^c, \nu^c$.
Then from the full Yukawa Lagrangian one can compute the charged fermion and Dirac neutrino mass matrices for the light fermions
written as $f^{c} M_f f$ as:

\vspace{-10pt}
\begin{equation}\label{eq:af}
M^{T}_{f} = v_{f} k_{f} \begin{bmatrix}
a^{f}_{11} & 0 & a^{f}_{13} \\
0 & a^{f}_{22} & a^{f}_{23} \\
a^{f}_{31} & a^{f}_{32} & a^{f}_{33}
\end{bmatrix} ,
\end{equation}

\noindent where $f=U, D, E, \nu_D$,   $v_{e}=v_{d}$, $v_{\nu}=v_{u}$, $k_{u,d}=1$ and $k_{e,\nu}=-3$. We define the ratio $v_{u}/v_{d} \equiv r$. 
Note that this ratio is not exactly equal to $\tan\beta$ of MSSM, but is closely related to it.  If we ignore the mixing of the up and down-type
Higgs doublets from $\overline{126}_H$ with other doublets present in the theory, $r$ would be equal to $\tan\beta$ in MSSM. 
The following relations are then readily obtained:

\vspace{-20pt}
\begin{align}
a^{f}_{11} &= a_{11} \; , \label{eq:af1} \\
a^{f}_{13}  &=   \frac{ a_{13}(e^{i \phi} \cos\theta + T Q_{f^{c}})-a_{14}\sin\theta }{N_{f^{c}}} \; ,\label{eq:af3}  \\
a^{f}_{22} &= a_{22} \; , \label{eq:af5} \\
a^{f}_{23} &=  \frac{a_{23} (e^{i \phi} \cos\theta + T Q_{f^{c}}) - a_{24}\sin\theta }{N_{f^{c}}}  \; , \label{eq:af6} \\
a^{f}_{31} &=  \frac{a_{13} (e^{i \phi} \cos\theta + T Q_{f}) - a_{14} \sin\theta }{N_{f}} \; , \label{eq:af7} \\
a^{f}_{32} &=   \frac{a_{23} (e^{i \phi} \cos\theta + T Q_{f}) - a_{24} \sin\theta }{N_{f}} \; , \label{eq:af8} \\
a^{f}_{33} &=   \frac{a_{33}(e^{i \phi} \cos\theta + T Q_{f}) (e^{i \phi} \cos\theta + T Q_{f^{c}}) +  a_{44}  \sin^{2}\theta - a_{34} \sin\theta [2 e^{i \phi} \cos\theta + T (Q_{f}+Q_{f^{c}})] }{N_{f} N_{f^{c}}}   \; \label{eq:af9}.
\end{align}

\noindent
Note that a rotation in the 1-2 sector has been made which makes $a^{f}_{12}=a^{f}_{21}=a_{12}=0$.
These mass matrices are not symmetric, since $a_{ij}^f \neq a_{ji}^f$, although the original matrix obeyes $a_{ij} = a_{ji}$.
These four mass matrices for $f=U,D,D,\nu_D$  are given in terms of the parameters $\epsilon ,T,\theta,\phi$ and $a_{ij}$ (with $i,j=1-4$
and $a_{12} = a_{21} = 0$).  We choose to take elements of $M_E$ to be independent.
One can then solve for $a_{13}$ and $a_{14}$ in terms of $a^{e}_{13}$ and $a^{e}_{31}$; similarly $a_{23}$ and $a_{24}$ in terms of $a^{e}_{23}$ and $a^{e}_{32}$. From Eqs.\eqref{eq:af3}, \eqref{eq:af6}, \eqref{eq:af7}, \eqref{eq:af8} and \eqref{eq:Q45} one  sees that this is a valid choice provided that $\epsilon \neq -5/9$ for $\Phi = 45_{H}$.
(If $\epsilon = -5/9$, $a^{e}_{13}=a^{e}_{31}$ and $a^{e}_{23}=a^{e}_{32}$, which does not lead to
realistic fermion masses.) Similarly for the case of $\Phi = 210_{H}$, the restriction is $\epsilon_{2} \neq 0$ is required as can be seen from Eq. \eqref{eq:Q210}. All these mass matrices have the same 1-2 sector  and one can choose $a_{11}=a^{e}_{11}$ and $ a_{22}=a^{e}_{22}$. In addition,  $a^{e}_{33},a^{u}_{33},a^{d}_{33}$ depend on 3 independent parameters  $ a_{33},a_{34},a_{44}$ that appear only in the (3,3) sector of
the light mass matrices.  Since this linear system is invertible, one can treat $a^{e}_{33},a^{u}_{33},a^{d}_{33}$ as independent parameters.
The (3,3) element of the right-handed neutrino Majorana matrix is then not free, and is determined in terms of $a^{e}_{33},a^{u}_{33},a^{d}_{33}$.
Expressions for $a_{ij}$ in terms of the independent parameters chosen are given in Appendix \ref{App:AppendixA} .

The elements of $M_{E}$ are independent parameters. We can express $M_{U}$ and $M_{D}$ in terms of $T, \theta, \phi, a^{u}_{33},a^{d}_{33}, a^{e}_{ij}$ and $\epsilon$ (or $\epsilon_{1,2}$) for the case of $45_{H}$ (or $210_{H}$), so in this basis the charged fermion mass matrices are:

\vspace{-12pt}
\begin{equation}\label{eq:lepton0}
M^{T}_{E} = -3 v_{d} \begin{bmatrix}
a^{e}_{11} & 0 & a^{e}_{13} \\
0 & a^{e}_{22} & a^{e}_{23} \\
a^{e}_{31} & a^{e}_{32} & a^{e}_{33}
\end{bmatrix} \; ;
\end{equation}

\vspace{-12pt}
\begin{equation}\label{eq:up}
M^{T}_{U}= v_{u} \begin{bmatrix}
a^{e}_{11} & 0 & \frac{a^{e}_{13} N_{e^{c}} (Q_{e}-Q_{u^{c}})+a^{e}_{31} N_{e} (-Q_{e^{c}}+Q_{u^{c}})}{N_{u^{c}} (Q_{e}-Q_{e^{c}}) } \\
0 & a^{e}_{22} &  \frac{a^{e}_{23} N_{e^{c}} (Q_{e}-Q_{u^{c}})+a^{e}_{32} N_{e} (-Q_{e^{c}}+Q_{u^{c}})}{N_{u^{c}} (Q_{e}-Q_{e^{c}}) } \\
\frac{a^{e}_{13} N_{e^{c}} (Q_{e}-Q_{u})+a^{e}_{31} N_{e} (-Q_{e^{c}}+Q_{u})}{N_{u} (Q_{e}-Q_{e^{c}}) } & \frac{a^{e}_{23} N_{e^{c}} (Q_{e}-Q_{u})+a^{e}_{32} N_{e} (-Q_{e^{c}}+Q_{u})}{N_{u} (Q_{e}-Q_{e^{c}}) } & a^{u}_{33}
\end{bmatrix} \; ;
\end{equation}

\vspace{-12pt}
\begin{equation}\label{eq:down}
M^{T}_{D}= v_{d} \begin{bmatrix}
a^{e}_{11} & 0 & \frac{a^{e}_{13} N_{e^{c}} (Q_{e}-Q_{d^{c}})+a^{e}_{31} N_{e} (-Q_{e^{c}}+Q_{d^{c}})}{N_{d^{c}} (Q_{e}-Q_{e^{c}}) } \\
0 & a^{e}_{22} &  \frac{a^{e}_{23} N_{e^{c}} (Q_{e}-Q_{d^{c}})+a^{e}_{32} N_{e} (-Q_{e^{c}}+Q_{d^{c}})}{N_{d^{c}} (Q_{e}-Q_{e^{c}}) } \\
\frac{a^{e}_{13} N_{e^{c}} (Q_{e}-Q_{u})+a^{e}_{31} N_{e} (-Q_{e^{c}}+Q_{u})}{N_{u} (Q_{e}-Q_{e^{c}}) } & \frac{a^{e}_{23} N_{e^{c}} (Q_{e}-Q_{u})+a^{e}_{32} N_{e} (-Q_{e^{c}}+Q_{u})}{N_{u} (Q_{e}-Q_{e^{c}}) } & a^{d}_{33}
\end{bmatrix} \; .
\end{equation}

\noindent
Since $Q_{u}=Q_{d}$, we have $a^{d}_{31}=a^{u}_{31}$ and $a^{d}_{32}=a^{u}_{32}$, see  Eqs. \eqref{eq:af7} and \eqref{eq:af8}.

Now, the rotation that was made in the 1-2 sector to set $a_{12}=0$ simultaneously can make $a^{e}_{11}$ and $a^{e}_{22}$ real. This rotation will alter the column $\{(M_{E})_{13}, (M_{E})_{23} \}^{T}$ and the row $\{(M_{E})_{31}, (M_{E})_{32} \}$ in such a way that the forms of $M_{U}$ Eq. \eqref{eq:up} and $M_{D}$ Eq. \eqref{eq:down} are preserved.  All parameters are complex, except that one among $a_{33}^{u,d,e}$ can be made real
(we choose $a_{33}^d$ to be real), and that $T$ can be chosen real.  So the parameter set is
\vspace{-8pt}
\begin{eqnarray*}
\{\epsilon, r, T, \theta, \phi, a^{e}_{11}, a^{e}_{22}, a^{e}_{13}, a^{e}_{31}, a^{e}_{23}, a^{e}_{32}, a^{e}_{33}, a^{u}_{33}, a^{d}_{33} \} \;\; \rm{for} \;\; 45_{H} \;\; \rm{or} \\
\{\epsilon_{1}, \epsilon_{2}, r, T, \theta, \phi, a^{e}_{11}, a^{e}_{22}, a^{e}_{13}, a^{e}_{31}, a^{e}_{23}, a^{e}_{32}, a^{e}_{33}, a^{u}_{33}, a^{d}_{33} \} \;\; \rm{for} \;\; 210_{H}.
\end{eqnarray*}
\noindent Of these sets, $\{a^{e}_{13},a^{e}_{31},a^{e}_{23},
a^{e}_{32},a^{e}_{33},a^{u}_{33} \}$ are complex (with $a^{d}_{33}$ chosen to be real). For $\Phi= 45_{H}$, there are 13 magnitudes and 7 phases (in total 20 parameters) for non-SUSY case. In the case of SUSY, $\epsilon$ is complex, so one additional phase enters (for a total 21 parameters).
For $\Phi= 210_{H}$ in the SUSY context with minimal Higgs content, $\epsilon_{1}$ and $\epsilon_2$ are not independent of each other (see later), so there are again 13 magnitudes and 8 phases (in total 21 parameters). Later we will also consider a case with non-minimal Higgs sector where both these VEV ratios $\epsilon_{1,2}$ can be in general independent of each
other. In the neutrino sector (discussed in the next subsection) the mass matrix is given by these same parameters except for an overall scale ($v_{R,L}$ for type-I and  type-II seesaw scenarios respectively) that adds one new parameter.

%%%%%%%%%%%%%%%%%%%%%%%%%%%%%%%%%%%%%%%
%%%%%%%%%%%%%%%%%%%%%%%%%%%%%%%%%%%%%%%
\subsection{The neutrino sector}
\subsubsection{Type-I seesaw}
\label{section:typeI}

To write down the mass matrix in the neutrino sector, we make the assumption that $M, b \Omega \gg v_{R}$, which is a valid approximation provided that $M, b \Omega \sim M_{GUT} \sim 10^{16}$ GeV.  Note that in order to generate light neutrino masses by using the seesaw mechanism, one roughly needs $v_{R} \sim 10^{12-14}$ GeV. In this approximation, no new parameter comes into play in the neutrino mass matrix except the scale $v_{R}$. For type-I seesaw mechanism the Dirac neutrino mass matrix can be read off from Eq. \eqref{eq:af}:

\vspace{-7pt}
\begin{equation}
M^{T}_{\nu_ D} = -3 v_{u} \begin{bmatrix}
a^{e}_{11} & 0 & \frac{a^{e}_{13} N_{e^{c}} (Q_{e}-Q_{\nu^{c}})+a^{e}_{31} N_{e} (-Q_{e^{c}}+Q_{\nu^{c}})}{N_{\nu^{c}} (Q_{e}-Q_{e^{c}}) } \\
0 & a^{e}_{22} &  \frac{a^{e}_{23} N_{e^{c}} (Q_{e}-Q_{\nu^{c}})+a^{e}_{32} N_{e} (-Q_{e^{c}}+Q_{\nu^{c}})}{N_{\nu^{c}} (Q_{e}-Q_{e^{c}}) } \\
a^{e}_{31} & a^{e}_{32} & a^{\nu}_{33}
\end{bmatrix}~.
\end{equation}

\noindent Since $Q_{e}=Q_{\nu}$, $a^{\nu}_{31}=a^{e}_{31}$ and $a^{\nu}_{32}=a^{e}_{32}$. The expression for $a^{\nu}_{33}$ are given in Eqs.  \eqref{anu3345} and \eqref{anu33210} for $\Phi=45_H$ and $210_H$ respectively in appendix \ref{App:AppendixB} . One can derive the right-handed neutrino Majorana mass matrix to be

\vspace{-12pt}
\begin{equation}
\frac{M_{\nu_ R}}{v_{R}}=   \begin{bmatrix}
a^{e}_{11} & 0 & \frac{a^{e}_{13} N_{e^{c}} (Q_{e}-Q_{\nu^{c}})+a^{e}_{31} N_{e} (-Q_{e^{c}}+Q_{\nu^{c}})}{N_{\nu^{c}} (Q_{e}-Q_{e^{c}}) } \\
0 & a^{e}_{22} &   \frac{a^{e}_{23} N_{e^{c}} (Q_{e}-Q_{\nu^{c}})+a^{e}_{32} N_{e} (-Q_{e^{c}}+Q_{\nu^{c}})}{N_{\nu^{c}} (Q_{e}-Q_{e^{c}}) }  \\
\frac{a^{e}_{13} N_{e^{c}} (Q_{e}-Q_{\nu^{c}})+a^{e}_{31} N_{e} (-Q_{e^{c}}+Q_{\nu^{c}})}{N_{\nu^{c}} (Q_{e}-Q_{e^{c}}) } &  \frac{a^{e}_{23} N_{e^{c}} (Q_{e}-Q_{\nu^{c}})+a^{e}_{32} N_{e} (-Q_{e^{c}}+Q_{\nu^{c}})}{N_{\nu^{c}} (Q_{e}-Q_{e^{c}}) } & a^{R}_{33}
\end{bmatrix},
\end{equation}
\noindent
The expressions for $a^{R}_{33}$ are given in Eqs.  \eqref{aR3345} and \eqref{aR33210} for $\Phi=45_H$ and $210_H$ respectively  in Appendix \ref{App:AppendixB}. Then, the light neutrino mass matrix in the type-I seesaw scenario is given  by
\vspace{-10pt}
\begin{equation}
M_{N} = - M_{\nu_ D}^T M^{-1}_{\nu_ R} M_{\nu_ D} .
\end{equation}

%%%%%%%%%%%%%%%%%%%%%%%%%%%%%%%%%%%%%%%
%%%%%%%%%%%%%%%%%%%%%%%%%%%%%%%%%%%%%%%
\subsubsection{Type-II seesaw}
\label{section:typeII}

In analogy to the the analysis done in Sec. \ref{section:typeI} one can  derive the type-II seesaw contributions to the the  neutrino mass matrix
by replacing $v_{R} \rightarrow v_{L}$ and $\nu^{c} \rightarrow \nu$. In this type-II seesaw scenario the neutrino mass matrix is then given by

\vspace{-12pt}
\begin{equation}
M_{\nu_L}  = v_{L} \begin{bmatrix}
a^{e}_{11} & 0 & a^{e}_{31} \\
0 & a^{e}_{22} & a^{e}_{32} \\
a^{e}_{31} & a^{e}_{32} & a^{L}_{33}.
\end{bmatrix}
\end{equation}
\noindent
The expressions for $a^{L}_{33}$ are given in Eqs.  \eqref{aL3345} and \eqref{aL33210} for $\Phi=45_H$ and $210_H$ respectively in Appendix \ref{App:AppendixB}.

%%%%%%%%%%%%%%%%%%%%%%%%%%%%%%%%%%%%%%%
%%%%%%%%%%%%%%%%%%%%%%%%%%%%%%%%%%%%%%%
\section{Symmetry breaking constraints}
\label{section:breaking}
In all models studied here, there is no $10_{H}$ Higgs and matter fields couple to $126_{H}+\overline{126}_{H}$ and $45_{H}$ or $210_{H}$ scalars.
There are considerations as outlined in Sec. II that would require additional Higgs fields to be present for consistent symmetry breaking.  While
there are no constraints on the VEV ratios when a $210_H$ is employed in the non-SUSY framework, these ratios are determined in the case of
SUSY.  We consider the various constraints on the symmetry breaking sector in this section.

%%%%%%%%%%%%%%%%%%%%%%%%%%%%%%%%%%%%%%%
%%%%%%%%%%%%%%%%%%%%%%%%%%%%%%%%%%%%%%%

\subsection{Non-SUSY \boldmath$SO(10)$ models A and B}

Model {\bf A} employs $126_H$, $45_H$ and a $54_H$.
Breaking of $SO(10)$ down to SM via $SU(5)$ channel is not viable due to gauge coupling unification and proton decay limits. If only $45_{H}$ and $126_{H}$ (or $16_H$) Higgs multiplets are used to break $SO(10)$, breaking takes place through the $SU(5)$-symmetric channel ~\cite{Yasue:1980fy, Barr:1981qv, Anastaze:1983zk}. The other two breaking channels $SO(10) \rightarrow SU(3)_{c}\times SU(2)_{L}\times SU(2)_{R}\times U(1)_{B-L} \rightarrow SM$ and $SO(10) \rightarrow SU(4)_{c}\times SU(2)_{L}\times U(1)_{R} \rightarrow SM$ do not have stable vacuum at the tree-level. Recently quantum corrections to the tree-level potential have been taken into account ~\cite{Bertolini:2009es, Bertolini:2010ng} and the validity of such breaking channels has been shown. However, we do not rely on quantum corrections in this paper. This is why the Higgs sector needs to be extended with a $54_H$ for consistent $SO(10)$ breaking down to SM ~\cite{Yasue:1980qj, Babu:1984mz}.  Note that a Higgs system consisting of $126_H$ and
$54_H$ is sufficient for symmetry breaking purposes if also a $10_H$ is used \cite{khan}, but without the $10_H$ as in our case, a $45_H$ is necessary.

Since the SM Higgs doublet is part of the $126_H$ in this model, a question arises as to the negativity of its squared mass.  Consistency of
the GUT scale symmetry breaking would require all physical scalar squared masses to be positive, which includes the SM Higgs doublet.  There must then
be a source that turns this positive mass to negative value.  It has been shown in Ref. \cite{nath} that indeed such a turn-around is possible,
provided that some scalar from any GUT multiplet remains light and has non-negligible couplings to the SM Higgs doublet.
The context in Ref. \cite{nath} is similar to our present case, where a $144_H$ of $SO(10)$ is used to break the GUT symmetry as well as the electroweak symmetry. Since our present non-SUSY model
has an intermediate scale, we expect some of the scalars to survive down to the intermediate scale, which would enable turning the Higgs mass-squared
to negative value so as to trigger electroweak symmetry breaking.

In Model {\bf B} we employ a $210_H$ in addition to the $126_H$. This is not however sufficient for our purpose. Since the VEV of $126_H$
is much smaller than the GUT scale, a single $210_H$ would break the GUT symmetry to one of its maximal little groups, such as $SU(5) \times U(1)$
or $SU(4)_c \times SU(2)_L \times SU(2)_R$ \cite{Chang:1985zq}.  The fermion mass matrices will then carry traces of this unbroken symmetry,
which would lead to unwanted mass relations.  This is why we extend the scalar sector by adding a $54_{H}$ or $16_{H}$. For non-SUSY $SO(10)$ model with Higgs multiplets $210_{H}+54_{H}$, since $54^{2}\ni 1_{s}+54_{s}+770_{s}$ and $210^{2}\ni 1_{s}+54_{s}+770_{s}$, the scalar potential contains 2 non-trivial quartic couplings between $210_{H}-54_{H}$. In addition, $210_{H}$ has 3 non-trivial quartic couplings and $54_{H}$ has one cubic and one non-trivial quartic couplings. This counting of non-trivial couplings dictates that in general the two VEV ratios $\epsilon_{1,2}$ from
the $210_H$ are free parameters. Similar argument can be provided if $54_{H}$ is replaced by $16_{H}$ Higgs.

\subsection{SUSY \boldmath$SO(10)$ Models C--F}

The Higgs sector of Model {\bf D} consists of $210_{H}+54_{H}+126_{H}+\overline{126}_{H}$. This system is a subset of the
 SUSY $SO(10)$ models studied in Ref. \cite{Fukuyama:2004ps}. The relevant part of the superpotential with only $210_{H}$, $54_{H}$ and $126_{H}+\overline{126}_{H}$ is:

\vspace{-15pt}
\begin{eqnarray}\label{eq:W}
\begin{aligned}
W &= \frac{1}{2} m_{1} \Phi^{2} + m_{2} \Delta \overline{\Delta} + \frac{1}{2} m_{5} E^{2} +\lambda_{1} \Phi^{3} + \lambda_{8} E^{3} \\ &+  \lambda_{2} \Phi \Delta \overline{\Delta}  + \lambda_{10} \Phi^{2} E + \lambda_{11} \Delta^{2} E  + \lambda_{12} \overline{\Delta}^{2} E .
\end{aligned}
\end{eqnarray}

Since the VEV of $126_{H}$ is required to be in the intermediate scale $\sim 10^{13-14}$ GeV range from a fit to light neutrino masses arising via the seesaw mechanism, in this analysis of the superpotential one can neglect the contribution coming from this field as the other scalars $210_{H}+54_{H}$ will get much larger VEVs of order the GUT scale $\sim 10^{16}$ GeV. Then the relevant stationary equations are

\vspace{-20pt}
\begin{eqnarray}
\begin{aligned}
0 &= m_{1} V_{1} + \frac{\lambda_{1}}{2 \sqrt{6}} V_{3}^{2} + \sqrt{\frac{3}{5}} V_{1} V_{E}, \\
0 &= m_{1} V_{2} + \frac{\lambda_{1}}{3 \sqrt{2}} (V_{2}^{2}+V_{3}^{2}) - \frac{2 \lambda_{10}}{\sqrt{15}} V_{2} V_{E}, \\
0 &= m_{1} V_{3} + \frac{\lambda_{1}}{\sqrt{6}} V_{1} V_{3}  + \frac{\sqrt{2} \lambda_{1}}{3} V_{2} V_{3} + \frac{\lambda_{10}}{2 \sqrt{15}} V_{3} V_{E}, \\
0 &= m_{5} V_{E} + \frac{\sqrt{3} \lambda_{8}}{2 \sqrt{5}} V_{E}^{2}  + \frac{\sqrt{3} \lambda_{10}}{2 \sqrt{5}} V_{1}^{2} - \frac{\lambda_{10}}{\sqrt{15}} V_{2}^{2} + \frac{\lambda_{10}}{4 \sqrt{15}} V_{3}^{2}.
\label{min}
\end{aligned}
\end{eqnarray}

\noindent
Here the $V_{1}= \langle (1,1,1) \rangle$, $V_{2}= \langle (1,1,15) \rangle$ and $V_{3}= \langle (1,3,15) \rangle$ are the VEVs
of $\Phi (210_H)$ and the $54_H$ VEV is $V_{E}= \langle (1,1,1) \rangle$ under the Pati-Salam group $SU(2)_{L}\times SU(2)_{R} \times SU(4)_{C}$ decomposition. Compared to Eqs. \eqref{eq:210vev}, here a different normalization is used and one can make the identifications $V_1=\phi_1, V_2=\sqrt{3} \phi_2, V_3=\sqrt{3/2} \phi_3$.

The last relation in Eq. (\ref{min}) can be solved for the free mass parameter $m_{5}$.  Taking differences of the other three twice, we
obtain two independent solutions,

\vspace{-10pt}
\begin{equation}
V_{1} = -\frac{\sqrt{3} V_{2} }{2} \;\;\;\rm{or,} \;\; V_{1} = \frac{ V^{2}_{3} }{2 \sqrt{3} V_{2}}.
\end{equation}
These correspond to the VEV ratios ratios ($\epsilon_{1}=V_{2}/V_{1}, \epsilon_{2}=V_{3}/V_{1}$) given as

\vspace{-20pt}
\begin{eqnarray}\label{eq:210A}
\epsilon_{1} = -\frac{2}{\sqrt{3}} \;\; \rm{or}, \;\; \epsilon_{1} = \frac{\epsilon_2^{2}}{2 \sqrt{3}}.
\end{eqnarray}

\noindent  While studing the fermions masses and mixing numerically, we will consider both these cases. These models are labelled as $\rm{\textbf{D}^{a}}$ for the solution $\epsilon_{1} = -\frac{2}{\sqrt{3}}$ and $\rm{\textbf{D}^{b}}$ for solution $\epsilon_{1} = \frac{\epsilon_2^{2}}{2 \sqrt{3}}$.

In Model {\bf E}, we use a $210_{H}$ along with a $16_{H} + \overline{16}_{H}$ for symmetry breaking purpose. These fields are in addition
to the $126_H+ \overline{126}_H$ fields present.
Just like the previous case, since the $SO(10)$ breaking VEV of the Higgs scalars $210_H$ and $16_{H} + \overline{16}_H$ are $\sim M_{GUT}$, one can neglect the terms involving the scalar $126_{H}$ which has a much lower VEV. The form of the superpotantial is identical to
Eq. \eqref{eq:W} with the $126_{H} (\overline{126}_{H})$ replaced by $16_{H} (\overline{16}_{H})$.  Denoting the $16_{H}(\overline{16}_{H})$ VEV as $V_{\psi}(\overline{V}_{\psi})$ and its mass by $m_{\psi}$, the relevant stationary equations in this case are

\vspace{-20pt}
\begin{eqnarray}
\begin{aligned}
0 &= m_{1} V_{1} + \frac{\lambda_{1}}{2 \sqrt{6}} V_{3}^{2} + \frac{\lambda_{2}}{10 \sqrt{6}} V_{\psi} \overline{V}_{\psi}, \\
0 &= m_{1} V_{2} + \frac{\lambda_{1}}{3 \sqrt{2}} (V_{2}^{2}+V_{3}^{2}) + \frac{\lambda_{2}}{10 \sqrt{2}} V_{\psi} \overline{V}_{\psi}, \\
0 &= m_{1} V_{3} + \frac{\lambda_{1}}{\sqrt{6}} V_{1} V_{3}  + \frac{\sqrt{2} \lambda_{1}}{3} V_{2} V_{3} + \frac{\lambda_{2}}{10} V_{\psi} \overline{V}_{\psi}, \\
0 &= V_{\psi} \overline{V}_{\psi} [m_{\psi}  + \frac{\lambda_{2}}{10 \sqrt{6}} V_{1} + \frac{\lambda_{2}}{10 \sqrt{2}} V_{2} + \frac{\lambda_{2}}{10} V_{3}].
\end{aligned}
\end{eqnarray}

There are two different solutions of this system of stationary equations

\vspace{-20pt}
\begin{eqnarray}
\begin{aligned}
V_{1}&= \frac{V_{3}}{\sqrt{6}},\;\; V_{2}= \frac{V_{3}}{\sqrt{2}} ; \\
\rm{or}, \;\; V_{1} &= \frac{-36 m^{2}_{1} V_{3}+5 V_{3}^{3} \lambda^{2}_{1}}{\sqrt{6} (-6 m_{1}+V_{3} \lambda_{1})^{2}} ,\;\;   V_{2} = -\frac{-36 m^{2}_{1} +12 m_{1}V_{3} \lambda_{1}+V^{2}_{3} \lambda^{2}_{1}}{\sqrt{2} \lambda_{1} (-6 m_{1}+V_{3} \lambda_{1})} .
\end{aligned}
\end{eqnarray}

\noindent
So the VEV ratios are given by

\vspace{-20pt}
\begin{eqnarray}\label{eq:210B}
\begin{aligned}
\epsilon_{1} &= \sqrt{3},\;\; \epsilon_{2} = \sqrt{6}; \\
\rm{or}, \;\; \epsilon_{1} &= \frac{\sqrt{3}(-6+\epsilon)(-36+12 \epsilon +\epsilon^{2})}{\epsilon (36-5 \epsilon^{2})} ,\;\; \epsilon_{2} = \frac{\sqrt{6}(-6+\epsilon)^{2}}{-36+5 \epsilon^{2}} \; ; \;\;\; \rm{with} \;\; \epsilon \equiv V_{3}\frac{\lambda_{1}}{m_{1}},
\end{aligned}
\end{eqnarray}

\noindent where $\epsilon$ is a free parameter. We discard the first solution since this corresponds to $SU(5)$-symmetric case.
The surviving model will be labeled {\bf E}.

By adding more Higgs multiplets in either of the models \textbf{D} or \textbf{E}, as for example $16_{H}+\overline{16}_{H}$ or adding another $54_{H}$ to model \textbf{D}, these relations for VEV ratios can be made invalid and $\epsilon_{1,2}$ can  be made independent parameters. We will also study this general case.  We choose to add $16_{H}+\overline{16}_{H}$ in model \textbf{D}  and $54_{H}$ in  in model \textbf{E} and label these classes of model as  \textbf{F}.  Finally, for SUSY model {\bf C}, consisting of $126_H+\overline{126}_H+45_H+54_H+16_H+\overline{16}_H$, we stress that
the $16_H+\overline{16}_H$ are needed for successfully tuning the MSSM doublets light without making simultaneously any other submultiplet light.
The parameter $\epsilon$ is arbitrary in this case.

%%%%%%%%%%%%%%%%%%%%%%%%%%%%%%%%%%%%%%%
%%%%%%%%%%%%%%%%%%%%%%%%%%%%%%%%%%%%%%%
\section{Numerical analysis of fermion masses and mixings}
In this section we show our fit results of fermion masses and mixings for different $SO(10)$ models described in Sections  II and V.
We do the fitting for both non-SUSY and SUSY cases, each with type-I and type-II seesaw scenarios. For optimization purpose we do
a $\chi^{2}$-analysis. The pull and $\chi^{2}$-function are defined as:
\vspace{-10pt}
\begin{align}
P_{i} &= \frac{O_{i\; \rm{th}}-E_{i\; \rm{exp}}}{\sigma_{i}}, \\
\chi^{2} &= \sum_{i} P_{i}^{2},
\end{align}

\noindent
where $\sigma_{i}$ represent experimental 1$\sigma$ uncertainty  and $O_{i\ \rm{th}}$, $E_{i\; \rm{exp}}$ and $P_{i}$ represent the theoretical prediction, experimental central value and pull of observable $i$. We fit the values of the observables at the GUT scale, $M_{GUT}=2\times 10^{16}$ GeV. To get the GUT scale values of the observables, for non-SUSY case, we take the central values at the $M_{Z}$ scale from Table-1 of Ref. \cite{Antusch:2013jca} and use the renormalization group equation (RGE) running factors given in Ref. \cite{Xing:2007fb} to get the GUT scale inputs. For the associated one sigma uncertainties of the observables at the GUT scale, we keep the same percentage uncertainty with respect to the central value of each quantity as that at the $M_{Z}$ scale. For SUSY case, the low scale values of the observables are taken from Table-2 of ~\cite{Antusch:2013jca} at $\mu=1$ TeV where the values are converted to the $\overline{\rm{DR}}$ scheme and  then using the renormalization group equation running for MSSM ~\cite{Barger:1992ac, Barger:1992pk} we get the GUT scale inputs. For all different SUSY $SO(10)$ models, we do the fitting for $\tan\beta=10$. For the charged lepton masses, a relative uncertainty of 0.1$\%$ is assumed in order to take into account the theoretical uncertainties arising for example from threshold effects. The inputs in the neutrino sector are taken from Ref. \cite{Fogli:2012ua}. For neutrino observables, we do not run the RGE from low scale to the GUT scale, which is a relatively small effect, except for an overall rescaling
on the neutrino masses that can be absorbed in the corresponding scale $v_{R}$ or $v_{L}$. In the case of inverted hierarchical neutrino mass spectrum, RGE effects can be important, whereas for all our cases the spectrum turns out to be normal hierarchical. Since the right-handed neutrino masses are extremely heavy, threshold corrections might also have effects on the neutrino observables if the Dirac neutrino matrix elements are of order one, but in our case the elements are much smaller than one.  All these inputs are shown in the tables where the fit results are presented. Below we present our best fit results and the corresponding parameters for different $SO(10)$ GUT models as discussed above.  \\

%\newpage
%%%%%%%%%%%%%%%%%%%%%%%%%%%%%%%%%%%%%%%
%%%%%%%%%%%%%%%%%%%%%%%%%%%%%%%%%%%%%%%
\noindent
\textbf{Model A: Non-SUSY} $\bm{SO(10)}$:  $\bm{45_{H}+54_{H}+\overline{126}_{H}}$

\FloatBarrier
\begin{table}[th]
\centering
\footnotesize
\resizebox{1.0\textwidth}{!}{
\begin{tabular}{|c|c|c|c|c|c|}
\hline
\pbox{10cm}{Masses (in GeV) and \\  Mixing parameters} & \pbox{10cm}{~~~~~Inputs \\ (at $\mu= M_{GUT}$)} & \pbox{10cm}{Fitted values (\textbf{AI})\\ (at $\mu= M_{GUT}$)} & \pbox{10cm}{pulls  \\ (\textbf{AI})} & \pbox{10cm}{Fitted values (\textbf{AII})\\ (at $\mu= M_{GUT}$)  } & \pbox{10cm}{pulls  \\ (\textbf{AII})}  \\ [1ex] \hline
$m_{u}/10^{-3}$ & 0.437$\pm$0.147 & 0.441 & 0.03&0.469 &0.21 \\ \hline
$m_{c}$   & 0.236$\pm$0.007 & 0.236 & 0.003&0.236 &0.02  \\ \hline
$m_{t}$   & 73.82$\pm$0.64 & 73.82 & 0.01&73.81 &-0.01 \\ \hline
$m_{d}/10^{-3}$  & 1.12$\pm$0.11 & 1.14 & 0.16&1.12 &-0.01 \\ \hline
$m_{s}/10^{-3}$  & 21.93$\pm$1.07 & 21.82 & -0.10&21.98 &0.04  \\ \hline
$m_{b}$  & 0.987$\pm$0.008 & 0.987 & -0.003&0.987 &-0.003 \\ \hline
$m_{e}/10^{-3}$   & 0.469658$\pm$0.000469 & 0.469649 & -0.01&0.469757 &0.21 \\ \hline
$m_{\mu}/10^{-3}$  & 99.1474$\pm$0.0991 & 99.1555 & 0.08&99.0913 &-0.56 \\ \hline
$m_{\tau}$  & 1.68551$\pm$0.00168 & 1.68542 & -0.05&1.68602 &0.29  \\ \hline
$|V_{us}|/10^{-2}$  & 22.54$\pm$0.06 & 22.53 & -0.01&22.54 &0.005 \\ \hline
$|V_{cb}|/10^{-2}$  & 4.856$\pm$0.06 & 4.856 & 0.001&4.853 &-0.03 \\ \hline
$|V_{ub}|/10^{-2}$  & 0.420$\pm$0.013 & 0.420 & 0.07&0.420 &0.02 \\ \hline
$\delta_{CKM}$ & 1.207$\pm$0.054  & 1.205 & -0.03&1.205 &-0.03 \\ \hline
$\Delta m^{2}_{sol}/10^{-5}$(eV$^{2}$) & 7.56$\pm$0.24 & 7.56 & 0.01&7.54 &-0.06 \\ \hline
$\Delta m^{2}_{atm}/10^{-3}$(eV$^{2}$) & 2.41$\pm$0.08 & 2.40 & -0.004&2.41 &0.05 \\ \hline
$\sin^{2}\theta^{PMNS}_{12}$ & 0.308$\pm$0.017 & 0.308 & 0.01&0.302 &-0.29  \\ \hline
$\sin^{2}\theta^{PMNS}_{23}$ & 0.387$\pm$0.0225 & 0.388 & 0.03&0.396 &0.42  \\ \hline
$\sin^{2}\theta^{PMNS}_{13}$ & 0.0241$\pm$0.0025 & 0.0238 & -0.11&0.0239 &-0.04  \\ \hline
\end{tabular}
}
\caption{ Fitted values of the observables correspond to $\chi^{2}=7\cdot10^{-2}$ and 0.78 for models \textbf{AI} and \textbf{AII} respectively. These fittings correspond to $|a_{ij}|_{max}=|a_{44}|=$1.9 and 3.3 for the type-I and type-II cases respectively (see text for details). For the charged lepton masses, a relative uncertainty of 0.1$\%$  is assumed in order to take into account the theoretical uncertainties arising for example from threshold effects. }
\label{tab:01}
\end{table}

%\FloatBarrier
\begin{table}[th]
\centering
\footnotesize
\resizebox{1.0\textwidth}{!}{
\begin{tabular}{|c|c|c|}
\hline
Quantity & \pbox{10cm}{Predicted Value (\textbf{AI})} & \pbox{10cm}{Predicted Value (\textbf{AII})}  \\ [1ex] \hline
$\{m_{1}, m_{2}, m_{3} \}$ (in eV) & $\{ 3.72\cdot10^{-3}, 9.45\cdot10^{-3},4.99\cdot10^{-2} \}$ & $\{ 4.38\cdot10^{-3}, 9.72\cdot10^{-3}, 5.00\cdot10^{-2} \}$ \\ \hline
$\{\delta^{PMNS}, \alpha^{PMNS}_{21}, \alpha^{PMNS}_{31} \}$ & $\{ 120.03^{\circ}, 144.92^{\circ}, -168.49^{\circ} \}$ & $\{ 104.80^{\circ}, 159.32^{\circ}, 95.05^{\circ} \}$ \\ \hline
$\{m_{cos}, m_{\beta}, m_{\beta \beta} \}$ (in eV) & $\{ 6.31\cdot10^{-2}, 6.55\cdot10^{-3}, 1.22\cdot10^{-3} \}$ & $\{ 6.42\cdot10^{-2}, 7.05\cdot10^{-3}, 2.24\cdot10^{-4} \}$ \\ \hline
$\{M_{1}, M_{2}, M_{3} \}$ (in GeV) & $\{ 8.65\cdot10^{7}, 2.66\cdot10^{10}, 6.99\cdot10^{11} \}$ & - \\ \hline
\end{tabular}
}
\caption{ Predictions of the models \textbf{A}. $m_{i}$ are the light neutrino masses, $M_{i}$ are the right-handed neutrino masses, $\alpha_{21,31}$ are the Majorana phases following the PDG parametrization, $m_{cos}=\sum_{i} m_{i}$, $m_{\beta}=\sum_{i} |U_{e i}|^{2} m_{i}$ is the effective mass parameter for beta-decay and $m_{\beta \beta}= | \sum_{i} U_{e i}^{2} m_{i} |$ is the effective mass parameter for neutrinoless double beta decay.}
\label{tab:02}
\end{table}

\noindent
The fit results and  the predictions for model \textbf{A} are shown in Table \ref{tab:01} and \ref{tab:02} respectively.
For model \textbf{AI} the parameter set is:
\vspace{-5pt}
\begin{align*}
\{ a^{u}_{33}, a^{d}_{33},  \epsilon, T, \theta, \phi, v_{R}, r \} = &\{ 0.415986 +0.0944114 i, 0.0246549, -1.24753,    8.68487, \\&  0.560999, -0.0127783, 1.58339\cdot10^{13} \rm{GeV},6.76689  \}
\end{align*}
\noindent and
\vspace{-10pt}
\begin{equation}
a^{e}_{ij} = 10^{-2} \left(
\begin{array}{ccc}
 0.0959072 & 0 & -1.47328-0.508307 i \\
 0 & -0.00693205 & -0.302045-0.119282 i \\
 0.149467\, +0.0128315 i & 0.0534903\, -0.0345252 i & 0.461306\, -1.4512 i \\
\end{array}
\right) .
\end{equation}

%%%%%%%%%%%%%%%%%%%%%%%%%%%%%%%%%%%%%%%%%%%

\noindent
For model \textbf{AII} (Model {\bf A} with type-II seesaw) the parameter set is:
\vspace{-5pt}
\begin{align*}
\{ a^{u}_{33}, a^{d}_{33},  \epsilon, T, \theta, \phi, v_{L}, r \}= &\{0.152744 +0.399269 i, -0.0244755, -0.393925, 11.4001, 0.560999, \\& 0.105066,  1.69937\cdot10^{-8} \rm{GeV}, 6.75824 \}
\end{align*}
\noindent and
\vspace{-10pt}
\begin{equation}
a^{e}_{ij} = 10^{-2} \left(
\begin{array}{ccc}
 0.127684 & 0 & -0.0742479+0.0532305 i \\
 0 & -0.00055042 & 0.0264824\, +0.0152045 i \\
 0.136072\, +0.0070994 i & 0.0582979\, +0.00164043 i &
   -0.398502-2.1619 i \\
\end{array}
\right) .
\end{equation}

In performing such optimization, solutions with lower values of $\chi^{2}$ exist but we are only interested in the solutions for which the original couplings $a_{ij}$ are also in the  perturbative range. In the optimization process we restrict ourselves to the case of $(a_{ij})_{max} \lesssim 2$. For all the solutions that are presented,  we did find good fits with this cut-off except for model \textbf{AII} where   $|a_{44}|= 3.3$ as can be seen from Eq. \ref{eq:aij} . The original coupling matrices $a_{ij}$ can be computed with the parameter sets that result due to the minimization process. For all the fits to the different models presented in this work, these matrices are shown in Appendix \ref{App:AppendixC}.  
%%%%%%%%%%%%%%%%%%%%%%%%%%%

\begin{figure}[th!]
\centering
\includegraphics[width=0.5\linewidth]{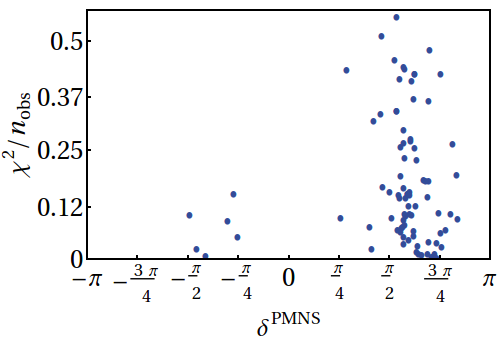}
\caption{ Variation of $\delta^{PMNS}$ with  $\chi^{2}/n_{obs}$ for the model \textbf{AI}.  In plotting this, we restrict to the regime for which $\chi^{2} \leq 10$. }\label{fig:del}
\end{figure}

In Table \ref{tab:02}, the predicted quantities correspond to the best fit values. For example, for model \textbf{AI}, the predicted value of the Dirac type CP violating phase in the neutrino sector is $\delta^{PMNS} = 2 \pi/3$. The fit result presented in this case is very good  since $\chi^2 =7\cdot 10^{-2}$. We have investigated the robustness of the predicted value of $\delta_{PMNS}$ and found it to be not very robust.
Sine the $\chi^2$  for the best fit is extremely small, it is quite fine to deviate from the minimum $\chi^2$ are still find acceptable fits.
We find that the variation of $\delta_{PMNS}$  from the predicted value can be quite large. In Fig. \ref{fig:del}, we show the variation of $\delta^{PMNS}$ with $\chi^{2}/n_{obs}$.  Most of the fit results presented in this work have small total $\chi^{2}$, so this conclusion on
the robustness of $\delta^{PMNS}$ prediction is valid for the other models as well.  We present the variation plot only for model \textbf{AI}.

%%%%%%%%%%%%%%%%%%%%%%%%%%%%%%%%%%%%%%%
%%%%%%%%%%%%%%%%%%%%%%%%%%%%%%%%%%%%%%%
\noindent
\textbf{Model B: Non-SUSY} $\bm{SO(10)}$:  $\bm{210_{H}+54_{H}+\overline{126}_{H}}$ (or $\bm{210_{H}+16_{H}+\overline{126}_{H}}$)

\noindent
The fit results and  the predictions for models \textbf{B} are shown in Table \ref{tab:03} and \ref{tab:04} respectively.
The parameter set for model \textbf{BI} is:
\vspace{-5pt}
\begin{align*}
\{ a^{u}_{33}, a^{d}_{33},  \epsilon_{1}, \epsilon_{2}, T, \theta, \phi, v_{R}, r \}=  &\{ -0.0380751-0.424441 i,  0.0244949, 1.61753, 1.67225, 0.764487, \\& 0.541654, -2.91319, 2.23915 \cdot10^{13} \rm{GeV}, 6.74578   \}
\end{align*}
\noindent and
\vspace{-10pt}
\begin{equation}
a^{e}_{ij} = 10^{-2} \left(
\begin{array}{ccc}
 -0.122115 & 0 & 0.899426\, +1.16951 i \\
 0 & 0.00569753 & -0.104101-0.15069 i \\
 -0.175821-0.103765 i & 0.0325028\, +0.0638096 i & 1.46544\, +0.663581 i
   \\
\end{array}
\right) .
\end{equation}

\noindent
The parameter set for model \textbf{BII} is:
\vspace{-5pt}
\begin{align*}
\{ a^{u}_{33}, a^{d}_{33},  \epsilon_{1}, \epsilon_{2}, T, \theta, \phi, v_{R}, r \}=  &\{ 0.174446\, +0.389832 i,  0.0244585, 1.07061, 0.666248,0.526787, \\& 0.713998,  0.295856,  1.32386 \cdot10^{-8} \rm{GeV}, 6.74635  \}
\end{align*}
\noindent and
\vspace{-10pt}
\begin{equation}
a^{e}_{ij} = 10^{-2} \left(
\begin{array}{ccc}
 0.00424453 & 0 & 0.130198\, -0.0532261 i \\
 0 & 0.0963929 & -0.386912-0.75915 i \\
 -0.0711623-0.0235054 i & 0.0531238\, +0.181145 i &
   -1.64856-1.16034 i \\
\end{array}
\right) .
\end{equation}

%\FloatBarrier
\begin{table}[th]
\centering
\footnotesize
\resizebox{1.0\textwidth}{!}{
\begin{tabular}{|c|c|c|c|c|c|}
\hline
\pbox{10cm}{Masses (in GeV) and \\  Mixing parameters} & \pbox{10cm}{~~~~~Inputs \\ (at $\mu= M_{GUT}$)} & \pbox{10cm}{Fitted values (\textbf{B I})\\ (at $\mu= M_{GUT}$)  } & \pbox{10cm}{pulls  \\ (\textbf{B I})} & \pbox{10cm}{Fitted values (\textbf{B II})\\ (at $\mu= M_{GUT}$)  } & \pbox{10cm}{pulls  \\ (\textbf{B II})}  \\ [1ex] \hline
$m_{u}/10^{-3}$ & 0.437$\pm$0.147 & 0.436 & -0.0007&0.437 &0.0002 \\ \hline
$m_{c}$   & 0.236$\pm$0.007 & 0.236 & 0.006&0.236 &-0.00009  \\ \hline
$m_{t}$   & 73.82$\pm$0.64 & 73.82 & 0.003&73.82 &-0.00005 \\ \hline
$m_{d}/10^{-3}$  & 1.12$\pm$0.11 & 1.12 & 0.0&1.12 &-0.0005 \\ \hline
$m_{s}/10^{-3}$  & 21.93$\pm$1.07 & 21.95 & 0.01&21.93 &-0.0003  \\ \hline
$m_{b}$  & 0.987$\pm$0.008 & 0.987 & 0.005&0.987 &0.0003 \\ \hline
$m_{e}/10^{-3}$   & 0.469658$\pm$0.000469 & 0.469654 & -0.008&0.469658 &-0.0004 \\ \hline
$m_{\mu}/10^{-3}$  & 99.1474$\pm$0.0991 & 99.1412 & -0.06&99.1476 &0.002 \\ \hline
$m_{\tau}$  & 1.68551$\pm$0.00168 & 1.68555 & 0.02&1.68551 &-0.002  \\ \hline
$|V_{us}|/10^{-2}$  & 22.54$\pm$0.06 & 22.54 & 0.0009&22.54 &-0.00004 \\ \hline
$|V_{cb}|/10^{-2}$  & 4.856$\pm$0.06 & 4.856 & 0.0001&4.856 &0.0002 \\ \hline
$|V_{ub}|/10^{-2}$  & 0.420$\pm$0.013 & 0.419 & -0.001&0.419 &-0.0001 \\ \hline
$\delta_{CKM}$ & 1.207$\pm$0.054  & 1.207 & 0.003&1.207 &0.0005 \\ \hline
$\Delta m^{2}_{sol}/10^{-5}$(eV$^{2}$) & 7.56$\pm$0.24 & 7.55 & -0.001&7.56 &0.00005 \\ \hline
$\Delta m^{2}_{atm}/10^{-3}$(eV$^{2}$) & 2.41$\pm$0.08 & 2.40 & 0.004&2.41 &0.0001 \\ \hline
$\sin^{2}\theta^{PMNS}_{12}$ & 0.308$\pm$0.017 & 0.307 & -0.004&0.307 &-0.0003  \\ \hline
$\sin^{2}\theta^{PMNS}_{23}$ & 0.387$\pm$0.0225 & 0.387 & -0.002&0.387 &0.00004  \\ \hline
$\sin^{2}\theta^{PMNS}_{13}$ & 0.0241$\pm$0.0025 & 0.0241 & 0.01&0.0241 &0.00009  \\ \hline
\end{tabular}
}
\caption{Best fit values of the observables correspond to $\chi^{2}=5\cdot10^{-3}$ and $1\cdot10^{-5}$ for models \textbf{BI} and \textbf{BII} respectively. These fittings correspond to $|a_{ij}|_{max}=|a_{44}|=$0.56 and 0.26 for the type-I and type-II cases respectively. For the charged lepton masses, a relative uncertainty of 0.1$\%$ is assumed in order to take into account the theoretical uncertainties arising for example from
threshold effects. }
\label{tab:03}
\end{table}

\FloatBarrier
\begin{table}[!htbp]
\centering
\footnotesize
\resizebox{1.0\textwidth}{!}{
\begin{tabular}{|c|c|c|}
\hline
Quantity & \pbox{10cm}{Predicted Value  (\textbf{BI})} & \pbox{10cm}{Predicted Value  (\textbf{BII})}   \\ [1ex] \hline
$\{m_{1}, m_{2}, m_{3} \}$ (in eV) & $\{ 2.58\cdot10^{-3}, 9.07\cdot10^{-3}, 4.99\cdot10^{-2} \}$  & $\{ 2.61\cdot10^{-3}, 9.07\cdot10^{-3}, 4.99\cdot10^{-2} \}$  \\ \hline

$\{\delta^{PMNS}, \alpha^{PMNS}_{21}, \alpha^{PMNS}_{31} \}$ & $\{ -38.38^{\circ}, 175.84^{\circ}, -131.48^{\circ} \}$ & $\{ -65.38^{\circ}, -158.28^{\circ}, -96.19^{\circ} \}$ \\ \hline

$\{m_{cos}, m_{\beta}, m_{\beta \beta} \}$ (in eV) & $\{ 6.15\cdot10^{-2}, 5.67\cdot10^{-3}, 8.33\cdot10^{-4} \}$ & $\{ 6.16\cdot10^{-2}, 5.69\cdot10^{-3}, 3.93\cdot10^{-4} \}$  \\ \hline

$\{M_{1}, M_{2}, M_{3} \}$ (in GeV) & $\{ 5.47\cdot10^{8}, 3.48\cdot10^{10}, 5.73\cdot10^{11} \}$ & - \\ \hline
\end{tabular}
}
\caption{ Predictions of models \textbf{B}. $m_{i}$ are the light neutrino masses, $M_{i}$ are the right handed neutrino masses, $\alpha_{21,31}$ are the Majorana phases following the PDG parametrization, $m_{cos}=\sum_{i} m_{i}$, $m_{\beta}=\sum_{i} |U_{e i}|^{2} m_{i}$ is the effective mass parameter for beta-decay and $m_{\beta \beta}= | \sum_{i} U_{e i}^{2} m_{i} |$ is the effective mass parameter for neutrinoless double beta decay.}
\label{tab:04}
\end{table}

%%%%%%%%%%%%%%%%%%%%%%%%%%%%%%%%%%%%%%%
%%%%%%%%%%%%%%%%%%%%%%%%%%%%%%%%%%%%%%%
\noindent
\textbf{Model C: SUSY} $\bm{SO(10)}$:  $\bm{45_{H}+54_{H}+16_{H} + \overline{16}_{H}+126_{H}+\overline{126}_{H}}$

\noindent
The fit results and  the predictions for models \textbf{C} are shown in Table \ref{tab:05} and \ref{tab:06} respectively.
The parameter set for model \textbf{CI} is:
\vspace{-5pt}
\begin{align*}
\{ a^{u}_{33}, a^{d}_{33},  \epsilon, T, \theta, \phi, v_{R}, r \}=  &\{ -0.22531+0.467722 i, 0.0317632, -1.10245+0.269791 i, \\& 5.51352, 0.560999,  0.448339, 1.39544\cdot 10^{13} \rm{GeV}, 9.14437 \}
\end{align*}
\noindent and
\vspace{-10pt}
\begin{equation}
a^{e}_{ij} = 10^{-2} \left(
\begin{array}{ccc}
 0.113952 & 0 & -1.52545-0.861024 i \\
 0 & 0.0066433 & 0.103718\, -0.25156 i \\
 0.103242\, -0.07705 i & 0.0528824\, +0.0275714 i & -0.300714-1.23388 i \\
\end{array}
\right) .
\end{equation}

\noindent
The parameter set for model \textbf{CII} is:
\vspace{-5pt}
\begin{align*}
\{ a^{u}_{33}, a^{d}_{33},  \epsilon, T, \theta, \phi, v_{L}, r \}=  &\{ 0.0307775 +0.518792 i, -0.0317378, -0.75526-0.237386 i, \\& 4.66699, 0.713998,  -0.0578946, 1.55237\cdot10^{-8} \rm{GeV}, 9.1424    \}
\end{align*}
\noindent and
\vspace{-10pt}
\begin{equation}
a^{e}_{ij} = 10^{-2} \left(
\begin{array}{ccc}
 -0.104302 & 0 & 0.88437\, -0.647303 i \\
 0 & 0.00422339 & 0.0628707\, -0.166004 i \\
 -0.0805026-0.140317 i & -0.0449238-0.0524454 i & -0.793384+1.69782 i \\
\end{array}
\right) .
\end{equation}

\FloatBarrier
\begin{table}[!htbp]
\centering
\footnotesize
\resizebox{1.0\textwidth}{!}{
\begin{tabular}{|c|c|c|c|c|c|}
\hline
\pbox{10cm}{Masses (in GeV) and \\  Mixing parameters} & \pbox{10cm}{~~~~~Inputs \\ (at $\mu= M_{GUT}$)} & \pbox{10cm}{Fitted values (\textbf{CI})\\ (at $\mu= M_{GUT}$) } & \pbox{10cm}{pulls  \\ (\textbf{CI})} & \pbox{10cm}{Fitted values (\textbf{CII})\\ (at $\mu= M_{GUT}$) } & \pbox{10cm}{pulls  \\ (\textbf{CII})}  \\ [1ex] \hline
$m_{u}/10^{-3}$ & 0.502$\pm$0.155 & 0.502 & 0.001&0.501 &-0.0005  \\ \hline
$m_{c}$   & 0.245$\pm$0.007 & 0.245 & 0.002&0.245 &0.001   \\ \hline
$m_{t}$   & 90.28$\pm$0.90 & 90.28 & -0.0005&90.28 &-0.002 \\ \hline
$m_{d}/10^{-3}$  & 0.839$\pm$0.084 & 0.838 & -0.006&0.839 &-0.001  \\ \hline
$m_{s}/10^{-3}$  & 16.62$\pm$0.90 & 16.62 & -0.00005&16.62 &0.002  \\ \hline
$m_{b}$  & 0.938$\pm$0.009 & 0.938 & -0.001&0.938 &-0.001  \\ \hline
$m_{e}/10^{-3}$   & 0.344021$\pm$0.000344 & 0.344021 & 0.0001&0.344018 &-0.008  \\ \hline
$m_{\mu}/10^{-3}$  & 72.6256$\pm$0.0726 & 72.6273 & 0.02&72.6240 &-0.02  \\ \hline
$m_{\tau}$  & 1.24038$\pm$0.00124 & 1.24036 & -0.01&1.24038 &-0.001  \\ \hline
$|V_{us}|/10^{-2}$  & 22.53$\pm$0.07 & 22.53 & 0.002 &22.53 &0.001 \\ \hline
$|V_{cb}|/10^{-2}$  & 3.934$\pm$0.06 & 3.933 & -0.001&3.934 &0.001   \\ \hline
$|V_{ub}|/10^{-2}$  & 0.340$\pm$0.011 & 0.340 & -0.001&0.340 &-0.004  \\ \hline
$\delta_{CKM}$ & 1.208$\pm$0.054  & 1.208 & 0.004 &1.208 &0.001  \\ \hline
$\Delta m^{2}_{sol}/10^{-5}$(eV$^{2}$) & 7.56$\pm$0.24 & 7.56 & 0.001&7.55 &-0.001  \\ \hline
$\Delta m^{2}_{atm}/10^{-3}$(eV$^{2}$) & 2.41$\pm$0.08 & 2.41 & 0.001&2.40 &-0.0006  \\ \hline
$\sin^{2}\theta^{PMNS}_{12}$ & 0.308$\pm$0.017 & 0.308 & 0.005&0.308 &0.003 \\ \hline
$\sin^{2}\theta^{PMNS}_{23}$ & 0.387$\pm$0.0225 & 0.387 & -0.0001 &0.387 &-0.002  \\ \hline
$\sin^{2}\theta^{PMNS}_{13}$ & 0.0241$\pm$0.0025 & 0.0240 & -0.002&0.0240 &-0.003    \\ \hline
\end{tabular}
}
\caption{Best fit result for models \textbf{C} with inputs correspond to $\tan\beta=10$. The fitted values correspond to $\chi^{2}=7\cdot10^{-4}$ for model \textbf{CI}  and $6\cdot10^{-4}$  for model \textbf{CII}. These fittings correspond to $|a_{ij}|_{max}=|a_{44}|=$1.5 and 1.03 for the type-I and type-II cases respectively.  For the charged lepton masses, a relative uncertainty of 0.1$\%$  is assumed in order to take into account the theoretical uncertainties arising for example from threshold effects. }
\label{tab:05}
\end{table}

\FloatBarrier
\begin{table}[!htbp]
\centering
\footnotesize
\resizebox{1.0\textwidth}{!}{
\begin{tabular}{|c|c|c|}
\hline
Quantity & \pbox{10cm}{Predicted Value  (\textbf{CI})} & \pbox{10cm}{Predicted Value  (\textbf{CII})}   \\ [1ex] \hline
$\{m_{1}, m_{2}, m_{3} \}$ (in eV) & $\{ 5.36\cdot10^{-3}, 1.02\cdot10^{-2}, 5.01\cdot10^{-2} \}$  & $\{ 3.68\cdot10^{-3}, 9.44\cdot10^{-3}, 4.99\cdot10^{-2} \}$  \\ \hline

$\{\delta^{PMNS}, \alpha^{PMNS}_{21}, \alpha^{PMNS}_{31} \}$ & $\{ 157.92^{\circ}, 158.41^{\circ}, -104.87^{\circ} \}$ & $\{ 85.15^{\circ}, 165.93^{\circ}, 138.22^{\circ} \}$ \\ \hline

$\{m_{cos}, m_{\beta}, m_{\beta \beta} \}$ (in eV) & $\{ 6.57\cdot10^{-2}, 7.90\cdot10^{-3}, 1.35\cdot10^{-3} \}$ & $\{ 6.31\cdot10^{-2}, 6.53\cdot10^{-3}, 7.55\cdot10^{-4} \}$  \\ \hline

$\{M_{1}, M_{2}, M_{3} \}$ (in GeV) & $\{ 1.91\cdot10^{8}, 1.63\cdot10^{10}, 1.33\cdot10^{12} \}$ & - \\ \hline
\end{tabular}
}
\caption{ Predictions of the models \textbf{C}. $m_{i}$ are the light neutrino masses, $M_{i}$ are the right handed neutrino masses, $\alpha_{21,31}$ are the Majorana phases following the PDG parametrization, $m_{cos}=\sum_{i} m_{i}$, $m_{\beta}=\sum_{i} |U_{e i}|^{2} m_{i}$ is the effective mass parameter for beta-decay and $m_{\beta \beta}= | \sum_{i} U_{e i}^{2} m_{i} |$ is the effective mass parameter for neutrinoless double beta decay.}
\label{tab:06}
\end{table}

\newpage
%%%%%%%%%%%%%%%%%%%%%%%%%%%%%%%%%%%%%%%
%%%%%%%%%%%%%%%%%%%%%%%%%%%%%%%%%%%%%%%
\noindent
\textbf{Model D: SUSY} $\bm{SO(10)}$:  $\bm{210_{H}+54_{H}+126_{H}+\overline{126}_{H}}$

\noindent
The fit results and  the predictions for model $\rm{\textbf{D}^{a}}$\textbf{I} are shown in Table \ref{tab:09A} and \ref{tab:10A} respectively. The parameter set for this fit of model $\rm{\textbf{D}^{a}}$\textbf{I} is:
\vspace{-5pt}
\begin{align*}
\{ a^{u}_{33}, a^{d}_{33},   \epsilon_2, T, \theta, \phi, v_{R}, r \}=  &\{ -0.343904+0.38917 i, 0.0318629, -5.89976-0.158839 i, 0.77736, \\& 0.532473, 2.76646, 1.95768 \cdot10^{13} \rm{GeV}, 9.15103       \}
\end{align*}
\noindent and
\vspace{-10pt}
\begin{equation}
a^{e}_{ij} =  10^{-10} \left(
\begin{array}{ccc}
 -0.00696426 & 0 & 0.289526\, -0.387539 i \\
 0 & -0.0703536 & 1.4192\, +0.447705 i \\
 0.00893467\, -0.0548221 i & 0.11474\, +0.140445 i &
   1.06975\, -1.07627 i \\
\end{array}
\right).
\end{equation}

%\FloatBarrier
\begin{table}[th]
\centering
\footnotesize
\resizebox{0.7\textwidth}{!}{
\begin{tabular}{|c|c|c|c|}
\hline
\pbox{10cm}{Masses (in GeV) and \\  Mixing parameters} & \pbox{10cm}{~~~~~Inputs \\ (at $\mu= M_{GUT}$)} & \pbox{10cm}{Fitted values ($\rm{\textbf{D}^{a}}$\textbf{I})\\ (at $\mu= M_{GUT}$) } & \pbox{10cm}{pulls  \\ ($\rm{\textbf{D}^{a}}$\textbf{I})}  \\ [1ex] \hline
$m_{u}/10^{-3}$ & 0.502$\pm$0.155  & 0.520 & 0.12 \\ \hline
$m_{c}$   & 0.245$\pm$0.007   & 0.243 & -0.20  \\ \hline
$m_{t}$   & 90.28$\pm$0.90  & 90.17 & -0.11  \\ \hline
$m_{d}/10^{-3}$  & 0.839$\pm$0.084  & 0.967 & 1.51   \\ \hline
$m_{s}/10^{-3}$  & 16.62$\pm$0.90  & 16.49 & -0.14  \\ \hline
$m_{b}$  & 0.938$\pm$0.009  & 0.939 & 0.14  \\ \hline
$m_{e}/10^{-3}$   & 0.344021$\pm$0.000344  & 0.343834 & -0.54 \\ \hline
$m_{\mu}/10^{-3}$  & 72.6256$\pm$0.0726  & 72.4978 & -1.75 \\ \hline
$m_{\tau}$  & 1.24038$\pm$0.00124  & 1.23997 & -0.32  \\ \hline
$|V_{us}|/10^{-2}$  & 22.53$\pm$0.07  & 22.53 & -0.09 \\ \hline
$|V_{cb}|/10^{-2}$  & 3.934$\pm$0.06  & 3.920 & -0.22   \\ \hline
$|V_{ub}|/10^{-2}$  & 0.340$\pm$0.011  & 0.341 & 0.10 \\ \hline
$\delta_{CKM}$ & 1.208$\pm$0.054   & 1.192 & -0.28 \\ \hline
$\Delta m^{2}_{sol}/10^{-5}$(eV$^{2}$) & 7.56$\pm$0.24   & 7.52 & -0.15 \\ \hline
$\Delta m^{2}_{atm}/10^{-3}$(eV$^{2}$) & 2.41$\pm$0.08  & 2.42 & 0.13  \\ \hline
$\sin^{2}\theta^{PMNS}_{12}$ & 0.308$\pm$0.017  & 0.290 & -1.00  \\ \hline
$\sin^{2}\theta^{PMNS}_{23}$ & 0.387$\pm$0.0225  & 0.399 & 0.55    \\ \hline
$\sin^{2}\theta^{PMNS}_{13}$ & 0.0241$\pm$0.0025  & 0.0235 & -0.20    \\ \hline
\end{tabular}
}
\caption{ Fitting result for model $\rm{\textbf{D}^{a}}$\textbf{I} with inputs correspond to $\tan\beta=10$. The fitted values correspond to $\chi^{2}=7.4$ for type-I. It should be mentioned that, among all the fit results presented in this work, this specific fit has the largest value of $\chi^{2}$ which is 7.4 for 18 observables.  This fit correspond to $|a_{ij}|_{max}=|a_{44}|=$1.55. For the charged lepton masses, a relative uncertainty of 0.1$\%$  is assumed in order to take into account theoretical uncertainties arising for example from threshold effects.   We did not find any acceptable fit within the perturbative range  for model $\rm{\textbf{D}^{a}}$\textbf{II}. }
\label{tab:09A}
\end{table}

%\FloatBarrier
\begin{table}[!htbp]
\centering
\footnotesize
\resizebox{0.7\textwidth}{!}{
\begin{tabular}{|c|c|}
\hline
Quantity & \pbox{10cm}{Predicted Value  ($\rm{\textbf{D}^{a}}$\textbf{I})}   \\ [1ex] \hline
$\{m_{1}, m_{2}, m_{3} \}$ (in eV)   & $\{ 1.58\cdot10^{-3}, 8.81\cdot10^{-3}, 4.99\cdot10^{-2} \}$    \\ \hline

$\{\delta^{PMNS}, \alpha^{PMNS}_{21}, \alpha^{PMNS}_{31} \}$  & $\{ 85.64^{\circ}, 139.76^{\circ}, 149.60^{\circ} \}$  \\ \hline

$\{m_{cos}, m_{\beta}, m_{\beta \beta} \}$ (in eV)  & $\{ 6.03\cdot10^{-2}, 4.78\cdot10^{-3}, 1.21\cdot10^{-3} \}$   \\ \hline

$\{M_{1}, M_{2}, M_{3} \}$ (in GeV)  & $\{ 8.89\cdot10^{7}, 2.14\cdot10^{10}, 2.63\cdot10^{12} \}$  \\ \hline
\end{tabular}
}
\caption{ Predictions of the model $\rm{\textbf{D}^{a}}$\textbf{I}. $m_{i}$ are the light neutrino masses, $M_{i}$ are the right handed neutrino masses, $\alpha_{21,31}$ are the Majorana phases following the PDG parametrization, $m_{cos}=\sum_{i} m_{i}$, $m_{\beta}=\sum_{i} |U_{e i}|^{2} m_{i}$ is the effective mass parameter for beta-decay and $m_{\beta \beta}= | \sum_{i} U_{e i}^{2} m_{i} |$ is the effective mass parameter for neutrinoless double beta decay.}
\label{tab:10A}
\end{table}

%%%%%%%%%%%%%%%%%%%%%%%%%%%%%%%%%%%%%%%%%
%%%%%%%%%%%%%%%%%%%%%%%%%%%%%%%%%%%%%%%%%
\noindent
The fit results and  the predictions for models $\rm{\textbf{D}^{b}}$ are shown in Table \ref{tab:09B} and \ref{tab:10B} respectively. The parameter set for $\rm{\textbf{D}^{b}}$\textbf{I} is:
\vspace{-5pt}
\begin{align*}
\{ a^{u}_{33}, a^{d}_{33},    \epsilon_{2}, T, \theta, \phi, v_{R}, r \}=  &\{-0.416619-0.310425 i , -0.0317247,  3.76592 +0.0145385 i, 0.310345, \\& 2.84818, 0.132797, 2.21257\cdot10^{13} \rm{GeV}, 9.14124 \}
\end{align*}
\noindent and
\vspace{-10pt}
\begin{equation}
a^{e}_{ij} = 10^{-2}  \left(
\begin{array}{ccc}
 0.0964978 & 0 & -0.0230964+1.18352 i \\
 0 & -0.00493562 & 0.00684639 -0.202567 i \\
 -0.0394903-0.200904 i & 0.055507 +0.0481135 i & 0.753644
   +1.64867 i
\end{array}
\right).
\end{equation}

%%%%%%%%%%%%%%%%%%%%%%%%%%%%%%%%%%%%%%%%%

\noindent
And  the parameter set for model $\rm{\textbf{D}^{b}}$\textbf{II} is:
\vspace{-5pt}
\begin{align*}
\{ a^{u}_{33}, a^{d}_{33},   \epsilon_{2}, T, \theta, \phi, v_{L}, r \}=  &\{ -0.365477-0.36971 i , 0.0316996, 3.53671 -0.311658 i,    0.343597, \\& -2.95969,     -0.131357,  1.58947\cdot10^{-8} \rm{GeV},  9.14446 \}
\end{align*}
\noindent and
\vspace{-10pt}
\begin{equation}
a^{e}_{ij} = 10^{-2} \left(
\begin{array}{ccc}
 0.00324628 & 0 & -0.026757+0.083972 i \\
 0 & -0.148375 & 0.450875 +0.843973 i \\
 -0.0190056-0.0577497 i & -0.129264-0.0587799 i & 1.86523
   +0.58344 i
\end{array}
\right) .
\end{equation}

%\FloatBarrier
\begin{table}[th]
\centering
\footnotesize
\resizebox{1.0\textwidth}{!}{
\begin{tabular}{|c|c|c|c|c|c|}
\hline
\pbox{10cm}{Masses (in GeV) and \\  Mixing parameters} & \pbox{10cm}{~~~~~Inputs \\ (at $\mu= M_{GUT}$)} & \pbox{10cm}{Fitted values ($\rm{\textbf{D}^{b}}$\textbf{I}) \\ (at $\mu= M_{GUT}$) } & \pbox{10cm}{pulls  \\ ($\rm{\textbf{D}^{b}}$\textbf{I})} & \pbox{10cm}{Fitted values ($\rm{\textbf{D}^{b}}$\textbf{II})\\ (at $\mu= M_{GUT}$) } & \pbox{10cm}{pulls  \\ ($\rm{\textbf{D}^{b}}$\textbf{II})}  \\ [1ex] \hline
$m_{u}/10^{-3}$ & 0.502$\pm$0.155  & 0.501 & -0.0006&0.502 &0.001  \\ \hline
$m_{c}$   & 0.245$\pm$0.007   & 0.245 & -0.004&0.245 &0.003  \\ \hline
$m_{t}$   & 90.28$\pm$0.90  & 90.28 & 0.002&90.28 &-0.00009  \\ \hline
$m_{d}/10^{-3}$  & 0.839$\pm$0.084  & 0.839 & 0.001&0.838 &-0.004   \\ \hline
$m_{s}/10^{-3}$  & 16.62$\pm$0.90  & 16.62 & -0.001&16.62 &-0.0001  \\ \hline
$m_{b}$  & 0.938$\pm$0.009  & 0.938 & -0.001&0.938 &0.001  \\ \hline
$m_{e}/10^{-3}$   & 0.344021$\pm$0.000344  & 0.344016 & -0.01&0.344019 &-0.007  \\ \hline
$m_{\mu}/10^{-3}$  & 72.6256$\pm$0.0726  & 72.6279 & 0.03&72.62249 &-0.01  \\ \hline
$m_{\tau}$  & 1.24038$\pm$0.00124  & 1.24035 & -0.02&1.24039 &0.004   \\ \hline
$|V_{us}|/10^{-2}$  & 22.53$\pm$0.07  & 22.53 & 0.0004&22.53 &-0.0003  \\ \hline
$|V_{cb}|/10^{-2}$  & 3.934$\pm$0.06  & 3.934 & 0.002 &3.933 &-0.0005  \\ \hline
$|V_{ub}|/10^{-2}$  & 0.340$\pm$0.011  & 0.340 & -0.001&0.340 &-0.0005  \\ \hline
$\delta_{CKM}$ & 1.208$\pm$0.054   & 1.208 & 0.002&1.208 &-0.001  \\ \hline
$\Delta m^{2}_{sol}/10^{-5}$(eV$^{2}$) & 7.55$\pm$0.24   & 7.56 & -0.0004&7.55 &-0.0003  \\ \hline
$\Delta m^{2}_{atm}/10^{-3}$(eV$^{2}$) & 2.41$\pm$0.08  & 2.41 & 0.0008&2.40 &-0.0003  \\ \hline
$\sin^{2}\theta^{PMNS}_{12}$ & 0.308$\pm$0.017  & 0.308 & -0.001&0.308 &-0.0003  \\ \hline
$\sin^{2}\theta^{PMNS}_{23}$ & 0.387$\pm$0.0225  & 0.387 & 0.0007 &0.387 &0.001    \\ \hline
$\sin^{2}\theta^{PMNS}_{13}$ & 0.0241$\pm$0.0025  & 0.0241 & 0.001 &0.02409 &-0.001   \\ \hline
\end{tabular}
}
\caption{ Fitting result for model $\rm{\textbf{D}^{b}}$ with inputs correspond to $\tan\beta=10$. The fitted values correspond to $\chi^{2}=1.9\cdot10^{-3}$ and $2\cdot10^{-4}$ for models $\rm{\textbf{D}^{b}}$\textbf{I} and $\rm{\textbf{D}^{b}}$\textbf{II} respectively.  These fits correspond to $|a_{ij}|_{max}=|a_{44}|=0.81$ and 0.99 for the two cases respectively. For the charged lepton masses, a relative uncertainty of 0.1$\%$  is assumed in order to take into account theoretical uncertainties arising for example from threshold effects. }
\label{tab:09B}
\end{table}

%\FloatBarrier
\begin{table}[!htbp]
\centering
\footnotesize
\resizebox{1.0\textwidth}{!}{
\begin{tabular}{|c|c|c|}
\hline
Quantity & \pbox{10cm}{Predicted Value  ($\rm{\textbf{D}^{b}}$\textbf{I})} & \pbox{10cm}{Predicted Value  ($\rm{\textbf{D}^{b}}$\textbf{II})}   \\ [1ex] \hline
$\{m_{1}, m_{2}, m_{3} \}$ (in eV)   & $\{ 2.20\cdot10^{-3}, 8.96\cdot10^{-3}, 4.99\cdot10^{-2} \}$ & $\{ 4.72\cdot10^{-3}, 9.89\cdot10^{-3}, 5.00\cdot10^{-2} \}$   \\ \hline

$\{\delta^{PMNS}, \alpha^{PMNS}_{21}, \alpha^{PMNS}_{31} \}$  & $\{ 50.24^{\circ}, 169.13^{\circ}, 111.61^{\circ} \}$  & $\{ 66.63^{\circ},161.63^{\circ}, 0.41^{\circ} \}$ \\ \hline

$\{m_{cos}, m_{\beta}, m_{\beta \beta} \}$ (in eV)  & $\{ 6.10\cdot10^{-2}, 5.38\cdot10^{-3}, 7.40\cdot10^{-4} \}$ & $\{ 6.47\cdot10^{-2}, 7.37\cdot10^{-3}, 4.54\cdot10^{-4} \}$  \\ \hline

$\{M_{1}, M_{2}, M_{3} \}$ (in GeV)  & $\{ 9.40\cdot10^{8}, 3.13\cdot10^{10}, 2.44\cdot10^{11} \}$ & - \\ \hline
\end{tabular}
}
\caption{ Predictions of models $\rm{\textbf{D}^{b}}$. $m_{i}$ are the light neutrino masses, $M_{i}$ are the right handed neutrino masses, $\alpha_{21,31}$ are the Majorana phases following the PDG parametrization, $m_{cos}=\sum_{i} m_{i}$, $m_{\beta}=\sum_{i} |U_{e i}|^{2} m_{i}$ is the effective mass parameter for  beta-decay and $m_{\beta \beta}= | \sum_{i} U_{e i}^{2} m_{i} |$ is the effective mass parameter for neutrinoless double beta decay.}
\label{tab:10B}
\end{table}

\newpage
%%%%%%%%%%%%%%%%%%%%%%%%%%%%%%%%%%%%%%%
%%%%%%%%%%%%%%%%%%%%%%%%%%%%%%%%%%%%%%%
\noindent
\textbf{Model E: SUSY} $\bm{SO(10)}$:  $\bm{210_{H}+16_{H} +\overline{16}_{H}+126_{H}+\overline{126}_{H}}$

\noindent
The fit results and  the predictions for models \textbf{E} are shown in Table \ref{tab:07} and \ref{tab:08} respectively.
\noindent
For model \textbf{EI}, the parameter set is:
\vspace{-5pt}
\begin{align*}
\{ a^{u}_{33}, a^{d}_{33}, \epsilon, T, \theta, \phi, v_{R}, r \}=  &\{ 0.0873809 +0.511807 i, 0.0316596, 3.21783 +0.31637 i, 0.762371, \\& 0.747998,  2.38528, 2.26917\cdot 10^{13} \rm{GeV}, 9.13917 \}
\end{align*}
\noindent and
\vspace{-10pt}
\begin{equation}
a^{e}_{ij} = 10^{-2} \left(
\begin{array}{ccc}
 0.00565532 & 0 & 0.0242668\, +0.230491 i \\
 0 & 0.10865 & 1.42287\, -0.445238 i \\
 -0.0636824-0.00136495 i & -0.154896+0.137372 i & -1.56252+0.079592 i
   \\
\end{array}
\right) .
\end{equation}

%%%%%%%%%%%%%%%%%%%%%
\noindent
For model \textbf{EII}, the parameter set is:
\vspace{-5pt}
\begin{align*}
\{ a^{u}_{33}, a^{d}_{33}, \epsilon, T, \theta, \phi, v_{L}, r \}= & \{ -0.43609+0.282193 i, -0.0316974, 3.21172 +0.154721 i, 0.54795, \\& 0.682955,  2.41863,  1.26299\cdot10^{-8} \rm{GeV}, 9.1442 \}
\end{align*}
\noindent and
\vspace{-8pt}
\begin{equation}
a^{e}_{ij} = 10^{-2} \left(
\begin{array}{ccc}
 -0.00491523 & 0 & 0.0991935\, +0.158945 i \\
 0 & -0.0989956 & 1.00507\, -0.775818 i \\
 -0.0677936-0.0254017 i & 0.0688011\, +0.186132 i & -1.33307+1.14438 i \\
\end{array}
\right).
\end{equation}

\FloatBarrier
\begin{table}[th]
\centering
\footnotesize
\resizebox{1.0\textwidth}{!}{
\begin{tabular}{|c|c|c|c|c|c|}
\hline
\pbox{10cm}{Masses (in GeV) and \\  Mixing parameters} & \pbox{10cm}{~~~~~Inputs \\ (at $\mu= M_{GUT}$)} & \pbox{10cm}{Fitted values (\textbf{EI})\\ (at $\mu= M_{GUT}$) } & \pbox{10cm}{pulls  \\ (\textbf{EI})} & \pbox{10cm}{Fitted values (\textbf{EII})\\ (at $\mu= M_{GUT}$) } & \pbox{10cm}{pulls  \\ (\textbf{EII})} \\ [1ex] \hline
$m_{u}/10^{-3}$ & 0.502$\pm$0.155  & 0.501 & -0.001&0.502 &0.0005  \\ \hline
$m_{c}$   & 0.245$\pm$0.007   & 0.245 & -0.007 &0.245 &0.001 \\ \hline
$m_{t}$   & 90.28$\pm$0.90  & 90.28 & 0.001&90.28 &-0.002  \\ \hline
$m_{d}/10^{-3}$  & 0.839$\pm$0.084  & 0.839 & 0.001 &0.839 &-0.0005  \\ \hline
$m_{s}/10^{-3}$  & 16.62$\pm$0.90  & 16.62 & 0.00009 &16.62 &-0.0001 \\ \hline
$m_{b}$  & 0.938$\pm$0.009  & 0.938 & -0.0002 &0.938 &-0.001 \\ \hline
$m_{e}/10^{-3}$   & 0.344021$\pm$0.000344  & 0.344022 & 0.004 &0.344023 &0.005 \\ \hline
$m_{\mu}/10^{-3}$  & 72.6256$\pm$0.0726  & 72.6250 & -0.007 &72.62641 &0.01 \\ \hline
$m_{\tau}$  & 1.24038$\pm$0.00124  & 1.24036 & -0.01  &1.24037 &-0.009  \\ \hline
$|V_{us}|/10^{-2}$  & 22.53$\pm$0.07  & 22.53 & 0.001 &22.53 &-0.0001 \\ \hline
$|V_{cb}|/10^{-2}$  & 3.934$\pm$0.06  & 3.934 & 0.005 &3.933 &-0.0003 \\ \hline
$|V_{ub}|/10^{-2}$  & 0.340$\pm$0.011  & 0.340 & -0.007 &0.340 &0.0006 \\ \hline
$\delta_{CKM}$ & 1.208$\pm$0.054   & 1.208 & 0.007&1.208 &0.004 \\ \hline
$\Delta m^{2}_{sol}/10^{-5}$(eV$^{2}$) & 7.56$\pm$0.24   & 7.56 & 0.001&7.55 &-0.0002 \\ \hline
$\Delta m^{2}_{atm}/10^{-3}$(eV$^{2}$) & 2.41$\pm$0.08  & 2.409 & -0.0007&2.41 &0.0003  \\ \hline
$\sin^{2}\theta^{PMNS}_{12}$ & 0.308$\pm$0.017  & 0.308 & 0.006&0.307 &-0.002 \\ \hline
$\sin^{2}\theta^{PMNS}_{23}$ & 0.387$\pm$0.0225  & 0.387 & 0.002&0.387 &0.0008     \\ \hline
$\sin^{2}\theta^{PMNS}_{13}$ & 0.0241$\pm$0.0025  & 0.0240 & 0.0001&0.0241 &0.001  \\ \hline
\end{tabular}
}
\caption{ Fitting result for models \textbf{E} with inputs correspond to $\tan\beta=10$. The fitted values correspond to $\chi^{2}=4\cdot10^{-4}$ for model \textbf{EI}  and $6\cdot10^{-4}$  for model \textbf{EII} respectively. These fittings correspond to $|a_{ij}|_{max}=|a_{44}|=$0.76 and 0.89 for the type-I and type-II cases respectively.  For the charged lepton masses, a relative uncertainty of 0.1$\%$  is assumed in order to take into account theoretical uncertainties arising for example from threshold effects.}
\label{tab:07}
\end{table}

\FloatBarrier
\begin{table}[th]
\centering
\footnotesize
\resizebox{1.0\textwidth}{!}{
\begin{tabular}{|c|c|c|}
\hline
Quantity & \pbox{10cm}{Predicted Value  (\textbf{EI})} & \pbox{10cm}{Predicted Value  (\textbf{EII})}   \\ [1ex] \hline
$\{m_{1}, m_{2}, m_{3} \}$ (in eV) & $\{ 2.06\cdot10^{-3}, 8.93\cdot10^{-3}, 4.98\cdot10^{-2} \}$  & $\{ 2.46\cdot10^{-3}, 9.03\cdot10^{-3}, 4.99\cdot10^{-2} \}$   \\ \hline

$\{\delta^{PMNS}, \alpha^{PMNS}_{21}, \alpha^{PMNS}_{31} \}$ & $\{ 46.84^{\circ}, -178.55^{\circ}, 141.46^{\circ} \}$ & $\{ -53.69^{\circ}, -172.46^{\circ}, -123.70^{\circ} \}$ \\ \hline

$\{m_{cos}, m_{\beta}, m_{\beta \beta} \}$ (in eV) & $\{ 6.08\cdot10^{-2}, 5.28\cdot10^{-3}, 9.54\cdot10^{-4} \}$  & $\{ 6.14\cdot10^{-2}, 5.58\cdot10^{-3}, 7.05\cdot10^{-4} \}$ \\ \hline

$\{M_{1}, M_{2}, M_{3} \}$ (in GeV)  & $\{ 2.79\cdot10^{8}, 2.15\cdot10^{10}, 1.82\cdot10^{12} \}$ & - \\ \hline
\end{tabular}
}
\caption{ Predictions of  models \textbf{E}. $m_{i}$ are the light neutrino masses, $M_{i}$ are the right handed neutrino masses, $\alpha_{21,31}$ are the Majorana phases following the PDG parametrization, $m_{cos}=\sum_{i} m_{i}$, $m_{\beta}=\sum_{i} |U_{e i}|^{2} m_{i}$ is the effective mass parameter for beta-decay and $m_{\beta \beta}= | \sum_{i} U_{e i}^{2} m_{i} |$ is the effective mass parameter for neutrinoless double beta decay.}
\label{tab:08}
\end{table}

%%%%%%%%%%%%%%%%%%%%%%%%%%%%%%%%%%%%%%%
%%%%%%%%%%%%%%%%%%%%%%%%%%%%%%%%%%%%%%%
\noindent
\textbf{Model F: SUSY} $\bm{SO(10)}$:  $\bm{210_{H}+54_{H}+16_H+\overline{16}_H+126_{H}+\overline{126}_{H}}$

\noindent
The fit results and  the predictions for models \textbf{F} are shown in Table \ref{tab:11} and \ref{tab:12} respectively. The parameter set for model \textbf{FI} is:
\vspace{-5pt}
\begin{align*}
\{ a^{u}_{33}, a^{d}_{33},  \epsilon_{1},  \epsilon_{2}, T, \theta, \phi, v_{R}, r \}=  &\{ -0.508413+0.106596 i, 0.0317542, 1.21369 +0.393457 i,  1.11752 \\&+1.12726 i, 0.652924, 0.682955, -2.69221,  2.17249\cdot10^{13} \rm{GeV}, 9.14433 \}
\end{align*}
\noindent and
\vspace{-10pt}
\begin{equation}
a^{e}_{ij} = 10^{-2} \left(
\begin{array}{ccc}
 -0.101322 & 0 & 1.50945\, +0.641937 i \\
 0 & 0.00628798 & -0.311418-0.017807 i \\
 -0.0659206-0.186996 i & -0.0254875+0.0490826 i & 0.979206\, +0.994616 i \\
\end{array}
\right) .
\end{equation}

%%%%%%%%%%%%%%%%%%%%%%%%%%%%%%%%%%%%%%%%%

\noindent
And the parameter set for model \textbf{FII} is:
\vspace{-5pt}
\begin{align*}
\{ a^{u}_{33}, a^{d}_{33},  \epsilon_{1},  \epsilon_{2}, T, \theta, &\phi, v_{L}, r \} =  \{ -0.0175831-0.518919 i, -0.0317748, 1.13488 -0.537296 i, \\& 0.934779 -0.810325 i,  0.577852, 0.541654, 2.37836, 1.17202\cdot10^{-8} \rm{GeV},  9.14329 \}
\end{align*}
\noindent and
\vspace{-10pt}
\begin{equation}
a^{e}_{ij} = 10^{-2} \left(
\begin{array}{ccc}
 0.00631171 & 0 & -0.244096-0.0355119 i \\
 0 & -0.106855 & 1.42499\, -0.0405503 i \\
 -0.00203514-0.0627216 i & -0.159398+0.138939 i & 1.56233\, +0.439752 i \\
\end{array}
\right) .
\end{equation}

\FloatBarrier
\begin{table}[th]
\centering
\footnotesize
\resizebox{1.0\textwidth}{!}{
\begin{tabular}{|c|c|c|}
\hline
Quantity & \pbox{10cm}{Predicted Value  (\textbf{FI})} & \pbox{10cm}{Predicted Value  (\textbf{FII})}   \\ [1ex] \hline
$\{m_{1}, m_{2}, m_{3} \}$ (in eV)   & $\{ 1.84\cdot10^{-3}, 8.88\cdot10^{-3}, 4.98\cdot10^{-2} \}$ & $\{ 2.00\cdot10^{-3}, 8.92\cdot10^{-3}, 4.98\cdot10^{-2} \}$   \\ \hline

$\{\delta^{PMNS}, \alpha^{PMNS}_{21}, \alpha^{PMNS}_{31} \}$  & $\{ -60.72^{\circ}, -175.43^{\circ}, -164.89^{\circ} \}$  & $\{ 44.97^{\circ},179.45^{\circ}, 133.12^{\circ} \}$ \\ \hline

$\{m_{cos}, m_{\beta}, m_{\beta \beta} \}$ (in eV)  & $\{ 6.06\cdot10^{-2}, 5.11\cdot10^{-3}, 1.17\cdot10^{-3} \}$ & $\{ 6.08\cdot10^{-2}, 5.23\cdot10^{-3}, 9.61\cdot10^{-4} \}$  \\ \hline

$\{M_{1}, M_{2}, M_{3} \}$ (in GeV)  & $\{ 3.92\cdot10^{8}, 1.97\cdot10^{10}, 1.27\cdot10^{12} \}$ & - \\ \hline
\end{tabular}
}
\caption{ Predictions of models \textbf{F}. $m_{i}$ are the light neutrino masses, $M_{i}$ are the right handed neutrino masses, $\alpha_{21,31}$ are the Majorana phases following the PDG parametrization, $m_{cos}=\sum_{i} m_{i}$, $m_{\beta}=\sum_{i} |U_{e i}|^{2} m_{i}$ is the effective mass parameter for beta-decay and $m_{\beta \beta}= | \sum_{i} U_{e i}^{2} m_{i} |$ is the effective mass parameter for neutrinoless double beta decay.}
\label{tab:12}
\end{table}

\FloatBarrier
\begin{table}[th]
\centering
\footnotesize
\resizebox{1.0\textwidth}{!}{
\begin{tabular}{|c|c|c|c|c|c|}
\hline
\pbox{10cm}{Masses (in GeV) and \\  Mixing parameters} & \pbox{10cm}{~~~~~Inputs \\ (at $\mu= M_{GUT}$)} & \pbox{10cm}{Fitted values (\textbf{FI})\\ (at $\mu= M_{GUT}$) } & \pbox{10cm}{pulls  \\ (\textbf{FI})} & \pbox{10cm}{Fitted values (\textbf{FII})\\ (at $\mu= M_{GUT}$) } & \pbox{10cm}{pulls  \\ (\textbf{FII})}  \\ [1ex] \hline
$m_{u}/10^{-3}$ & 0.502$\pm$0.155  & 0.501 & -0.003&0.501 &-0.0005  \\ \hline
$m_{c}$   & 0.245$\pm$0.007   & 0.245 & 0.006&0.245 &0.001  \\ \hline
$m_{t}$   & 90.28$\pm$0.90  & 90.28 & 0.003&90.28 &0.001  \\ \hline
$m_{d}/10^{-3}$  & 0.839$\pm$0.084  & 0.839 & 0.004&0.839 &0.001   \\ \hline
$m_{s}/10^{-3}$  & 16.62$\pm$0.90  & 16.62 & -0.001&16.62 &0.001  \\ \hline
$m_{b}$  & 0.938$\pm$0.009  & 0.938 & 0.0001&0.938 &-0.0001  \\ \hline
$m_{e}/10^{-3}$   & 0.344021$\pm$0.000344  & 0.344022 & 0.001&0.344022 &0.002  \\ \hline
$m_{\mu}/10^{-3}$  & 72.6256$\pm$0.0726  & 72.6237 & -0.02&72.62539 &-0.002  \\ \hline
$m_{\tau}$  & 1.24038$\pm$0.00124  & 1.24039 & 0.007&1.24038 &0.0003   \\ \hline
$|V_{us}|/10^{-2}$  & 22.53$\pm$0.07  & 22.53 & 0.0002&22.53 &0.0001  \\ \hline
$|V_{cb}|/10^{-2}$  & 3.934$\pm$0.06  & 3.933 & -0.001 &3.934 &0.0001  \\ \hline
$|V_{ub}|/10^{-2}$  & 0.340$\pm$0.011  & 0.340 & -0.007&0.340 &-0.001  \\ \hline
$\delta_{CKM}$ & 1.208$\pm$0.054   & 1.208 & 0.001&1.208 &0.004  \\ \hline
$\Delta m^{2}_{sol}/10^{-5}$(eV$^{2}$) & 7.56$\pm$0.24   & 7.56 & 0.00003&7.55 &-0.0002  \\ \hline
$\Delta m^{2}_{atm}/10^{-3}$(eV$^{2}$) & 2.41$\pm$0.08  & 2.41 & 0.0005&2.41 &0.0001  \\ \hline
$\sin^{2}\theta^{PMNS}_{12}$ & 0.308$\pm$0.017  & 0.308 & 0.0004&0.308 &0.0004  \\ \hline
$\sin^{2}\theta^{PMNS}_{23}$ & 0.387$\pm$0.0225  & 0.387 & -0.001 &0.387 &0.001    \\ \hline
$\sin^{2}\theta^{PMNS}_{13}$ & 0.0241$\pm$0.0025  & 0.0240 & -0.0009 &0.02409 &-0.002   \\ \hline
\end{tabular}
}
\caption{ Fitting result for models \textbf{F} with inputs correspond to $\tan\beta=10$. The fitted values correspond to $\chi^{2}=9\cdot10^{-4}$   and $3\cdot10^{-5}$  for model \textbf{FI}  and $6\cdot10^{-4}$  for model \textbf{FII} respectively. These fittings correspond to $|a_{ij}|_{max}=|a_{44}|=$0.67 and 1.08 for the type-I and type-II cases respectively.  For the charged lepton masses, a relative uncertainty of 0.1$\%$  is assumed in order to take into account theoretical uncertainties arising for example from threshold effects. }
\label{tab:11}
\end{table}

\vspace*{-0.2in}

%%%%%%%%%%%%%%%%%%%%%%%%%%%%%%%%%%%%%%%%
%%%%%%%%%%%%%%%%%%%%%%%%%%%%%%%%%%%%%%%%
%%%%%%%%%%%%%%%%%%%%%%%%%%%%%%%%%%%%%%%%
\section{$d=5$ proton decay}

Since the flavor dynamics occurs at the GUT scale in this class of models, the best hope for testing this idea
is by studying proton decay, in particular, its branching ratios into different modes.  While such an analysis can
be done for both non-SUSY and SUSY models, here we confine our discussion to the more dominant $d=5$ decay modes
in SUSY mediated by the color-triplet Higgsinos.

We will bound ourselves to the (presumably) dominant $d=5$ (charged) wino
mediated mode, so that only SU(2)$_L$ non-singlets will appear in the
effective operators:
\vspace{-5pt}
\beq
W\propto
\left(Y_{QQ}\right)_{ij}\left(Y_{QL}\right)_{kl}\left(Q_iQ_j\right)\left(Q_kL_l\right)
\eeq

\noi
with
\vspace{-9pt}
\begin{align}
\label{YQQ}
Y_{QQ}&=\Lambda_Q^T\left(Y-yx_Q^T-x_Qy^T+y_4x_Qx_Q^T\right)\Lambda_Q\\
\label{YQL}
Y_{QL}&=\Lambda_Q^T\left(Y-yx_L^T-x_Qy^T+y_4x_Qx_L^T\right)\Lambda_L
\end{align}

We have to project them to the mass eigenstates defined by the unitary matrices
$X = U, D, E, N$ which diagonalize the mass matrices as
\vspace{-5pt}
\beq
M_X=X_RM_X^dX_L^\dagger
\eeq

We will use the notation ($X=U,D$)
\vspace{-5pt}
\bea
Y_{XZ}&=&X_L^TY_{QQ}Z_L\;\;({\rm for}\;\;Z=U,D)\\
&=&X_L^TY_{QL}Z_L\;\;({\rm for}\;\;Z=E,N)
\eea

After 1-loop $\tilde w^\pm$ dressing
and assuming degeneracy and negligible left-right sfermion mixing the normalized amplitudes for different
channels \cite{Babu:1995cw} are, in the mass eigenbasis,
\vspace{-3pt}
\begin{align}
\label{Kplusnu}
A(K^+\bar\nu_l)&=\langle K^+|\left(ud\right)_Ls_L|p\rangle
\left[\left(Y_{UD}\right)_{11}\left(Y_{DN}\right)_{2l}-
\left(Y_{DD}\right)_{21}\left(Y_{UN}\right)_{1l}\right]\non\\
&+\langle K^+|\left(us\right)_Ld_L|p\rangle
\left[\left(Y_{UD}\right)_{12}\left(Y_{DN}\right)_{1l}-
\left(Y_{DD}\right)_{12}\left(Y_{UN}\right)_{1l}\right]\\
A(\pi^+\bar\nu_l)&=\langle \pi^+|\left(ud\right)_Ld_L|p\rangle
\left[\left(Y_{UD}\right)_{11}\left(Y_{DN}\right)_{1l}-
\left(Y_{DD}\right)_{11}\left(Y_{UN}\right)_{1l}\right]\\
A(K^0e_l^+)&=\langle K^0|\left(us\right)_Lu_L|p\rangle
\left[\left(Y_{UU}\right)_{11}\left(Y_{DE}\right)_{2l}-
\left(Y_{UD}\right)_{12}\left(Y_{UE}\right)_{1l}\right]\\
A(\pi^0e_l^+)&=\langle \pi^0|\left(ud\right)_Lu_L|p\rangle
\left[\left(Y_{UU}\right)_{11}\left(Y_{DE}\right)_{1l}-
\left(Y_{UD}\right)_{11}\left(Y_{UE}\right)_{1l}\right]\\
\label{etae}
A(\eta e_l^+)&=\langle \eta|\left(ud\right)_Lu_L|p\rangle
\left[\left(Y_{UU}\right)_{11}\left(Y_{DE}\right)_{1l}-
\left(Y_{UD}\right)_{11}\left(Y_{UE}\right)_{1l}\right]
\end{align}

\noi
where the numerical values (with maximal error around $30\%$) of the hadron matrix
elements can be found in \cite{Aoki:2013yxa}.

The unitary matrices $X$ and the Yukawa matrix elements $Y_{QQ,QL}$ are outputs of each successful fit done.
As an example, for model $\rm{\textbf{D}^a}$\textbf{I} we find
\vspace{-5pt}
\begin{align}
Y_{QQ}&=
\left(
\begin{array}{ccc}
 -0.0000696426 & 0 & -0.0105713+0.00524935 i \\
 0 & -0.000703536 & -0.0237115-0.0274144 i \\
 -0.0105713+0.00524935 i & -0.0237115-0.0274144 i & -1.05171-0.204611 i \\
\end{array}
\right)\\
Y_{QL}&=
\left(
\begin{array}{ccc}
 -0.0000696426 & 0 & -0.0000232394-0.000554968 i \\
 0 & -0.000703536 & 0.00140745\, +0.00114372 i \\
 -0.0105713+0.00524935 i & -0.0237115-0.0274144 i & 0.00550524\, -0.00420826 i \\
\end{array}
\right)\\
U_{L}&=
\left(
\begin{array}{ccc}
 0.947932\, +0.154511 i & 0.0250533\, -0.277159 i & -0.00483953-0.00916537 i \\
 0.0236485\, +0.277423 i & -0.948101+0.150221 i & -0.0314764+0.00505612 i \\
 -0.00438488-0.00319527 i & 0.0288651\, +0.0161593 i & -0.895175-0.444452 i \\
\end{array}
\right)\\
D_{L}&=
\left(
\begin{array}{ccc}
 0.44376\, +0.785783 i & -0.114306-0.415343 i & -0.00683343+0.0000682437 i \\
 -0.135958+0.408742 i & -0.484535+0.761308 i & 0.00448262\, -0.00785931 i \\
 -0.00402935-0.00717747 i & -0.00772054-0.00109752 i & -0.895597-0.444722 i \\
\end{array}
\right)\\
E_{L}&=
\left(
\begin{array}{ccc}
 -0.914868+0.192948 i & -0.182083-0.209497 i & -0.00368156+0.220751 i \\
 0.16774\, -0.285354 i & -0.612359-0.228437 i & 0.639383\, +0.233362 i \\
 -0.0189384-0.125958 i & -0.00319539+0.704116 i & -0.0568573+0.696242 i \\
\end{array}
\right)\\
N_{L}&=
\left(
\begin{array}{ccc}
 -0.502397+0.721475 i & -0.139083-0.437348 i & 0.047407\, +0.119192 i \\
 -0.43546+0.122682 i & -0.243261+0.682259 i & 0.24818\, -0.45725 i \\
 -0.0953668+0.115327 i & 0.0333823\, +0.513435 i & -0.0502508+0.842822 i \\
\end{array}
\right)
\end{align}

After squaring (\ref{Kplusnu})-(\ref{etae}) and multiplying by the appropriate phase space factor
($m_P$, $m_L$,  $m_p$ are the pseudo-scalar, lepton and proton mass, respectively)
\vspace{-5pt}
\beq
\left(1-2\left(\frac{m_P^2+m_L^2}{m_p^2}\right)+\left(\frac{m_P^2-m_L^2}{m_p^2}\right)^2\right)
\eeq

\noi
one can calculate the branching fractions for different channels (for neutrino final states we sum
over all 3 flavors), the results are given for the different models in table \ref{pdk}.  While as expected,
the $K^+ \overline{\nu}$ mode dominates, other sub-leading modes, notably $p \rightarrow \pi^+ \overline{\nu}$,
can be used to test and distinguish various models.

\begin{table}[h!]
  \centering
  \begin{tabular}{|c||r|r|r|r|r|r|r|r|r|}
   \hline
 & \multicolumn{1}{|c|}{\textbf{CI}} & \multicolumn{1}{|c|}{\textbf{CII}} & \multicolumn{1}{|c|}{$\rm{\textbf{D}^a}$ \textbf{I}}
 & \multicolumn{1}{|c|}{$\rm{\textbf{D}^b}$\textbf{I}} & \multicolumn{1}{|c|}{$\rm{\textbf{D}^b}$\textbf{II}} & \multicolumn{1}{|c|}{\textbf{EI}}
 & \multicolumn{1}{|c|}{\textbf{EII}} & \multicolumn{1}{|c|}{\textbf{FI}} & \multicolumn{1}{|c|}{\textbf{FII}}\\ \hline \hline
$K^+\bar \nu$ &  88.39 & 94.36 & 50.39 & 92.71 & 75.26 & 89.03 & 77.91 & 94.78 & 90.65 \\ \hline
$\pi^+\bar\nu$ & 10.85 & 5.55 & 48.33 & 7.12 & 24.62 & 10.48 & 21.58 & 4.95 & 9.17 \\ \hline
$ K^0 e^+$ & 0.00 & 0.00 & 0.00 & 0.00 & 0.00 & 0.00 & 0.00 & 0.00 & 0.00 \\ \hline
$K^0 \mu^+$ & 0.35 & 0.04 & 0.49 & 0.08 & 0.05 & 0.23 & 0.21 & 0.13 & 0.09 \\ \hline
$\pi^0 e^+$ & 0.00 & 0.00 & 0.00 & 0.00 & 0.00 & 0.00 & 0.00 & 0.00 & 0.00 \\ \hline
$\pi^0 \mu^+$ & 0.34 & 0.04 & 0.66 & 0.08 & 0.06 & 0.21 & 0.25 & 0.12 & 0.08 \\ \hline
$\eta e^+$ &  0.00 & 0.00 & 0.00 & 0.00 & 0.00 & 0.00 & 0.00 & 0.00 & 0.00 \\ \hline
$\eta \mu^+$ & 0.06 & 0.01 & 0.12 & 0.01 & 0.01 & 0.04 & 0.05 & 0.02 & 0.01 \\ \hline
  \end{tabular}
  \caption{Branching ratios for the main decay modes of the proton mediated by colored Higgsinos in SUSY $SO(10)$ models
  with successful fermion fits.}
  \label{pdk}
\end{table}

%%%%%%%%%%%%%%%%%%%%%%%%%%%%%%%%%%%%%%%
%%%%%%%%%%%%%%%%%%%%%%%%%%%%%%%%%%%%%%%
\section{Conclusion}

We have presented in this paper a new class of $SO(10)$ models that can successfully address the flavor puzzle.
The key ingredient of our models is the absence of $10_H$ that is conventionally used in most $SO(10)$ models.
Its absence is compensated by the introduction of a vector-like family in the $16+\overline{16}$ representation.
The Yukawa sector of these models has just a single $4 \times 4$ matrix, along with two four-vectors. As a consequence,
there are only 14 flavor parameters and 7 phases to fit all fermion masses and mixings, including the neutrino sector.

While the Yukawa system is highly nonlinear, by numerical optimization we have found excellent fits to the fermion
observables in a variety of models.  A $\overline{126}_H$ is present in all models, to generate large right-handed
neutrino Majorana masses as well as to provide the SM Higgs doublet.  The vector-like fermions have couplings to either
a $45_H$ or a $210_H$ that is used to complete the symmetry breaking. A total of six models, four supersymmetric and
two non-supersymmetric, have been studied.  In each case type-I or type-II seesaw mechanism was analyzed.  In one case
(Model {\bf D}) with SUSY, minimization of the Higgs potential led to a two-fold solution set, with each providing
an excellent fit to flavor observables.

While this class of high scale models cannot be easily tested at collider experiments, proton decay provides an avenue
to probe such models. We have investigate the branching ratios for proton decay in the SUSY models, with the results
presented in Table \ref{pdk}.  While it is an ambitious goal to test flavor models in proton decay discovery, even without
such a discovery it is heartening to learn that a large class of models can shed light on the various puzzles of fermion
masses observed in nature. In particular, starting from a highly symmetrical quark and lepton sector these models produce
large neutrino mixing simultaneously with small quark mixing, a highly nontrivial achievement.

%%%%%%%%%%%%%%%%%%%%%%%%%%%%%%%%%%%%%%%%%%%%%
%%%%%%%%%%%%%%%%%%%%%%%%%%%%%%%%%%%%%%%%%%%%%
\section*{Acknowledgments}
This work is supported in part by the U.S. Department of Energy Grant No. de-sc0010108 (K.S.B and S.S).
The work of B.B. is supported by the Slovenian Research Agency. Numerical calculations are performed  with
Cowboy Supercomputer of the High Performance Computing Center at Oklahoma State University (NSF grant
no. OCI-1126330). The authors wish to thank Barbara Szczerbinska and the CETUP* 2015 workshop for hospitality
and for providing a stimulating environment for discussions.

%%%%%%%%%%%%%%%%%%%%%%%%%%%%%%%%%%%%%%%
%%%%%%%%%%%%%%%%%%%%%%%%%%%%%%%%%%%%%%%
\begin{appendices}
\appendixpageoff

\section{Expressions for $a_{ij}$} \label{App:AppendixA}
In this Appendix we give expressions for $a_{ij}$ used in the numerical analysis.

\begin{align}
a_{13} &= \frac{N_{e^{c}} a^{e}_{13} - N_{e} a^{e}_{31}}{T (Q_{e^{c}}-Q_{e})} \; , \\
a_{14} &= \frac{N_{e^{c}} a^{e}_{13} (e^{i \phi} \cos\theta + T Q_{e})- N_{e} a^{e}_{31} (e^{i \phi} \cos\theta + T Q_{e^{c}}) }{\sin\theta T (Q_{e^{c}}-Q_{e})} \; , \\
a_{23} &=  \frac{N_{e^{c}} a^{e}_{23} - N_{e} a^{e}_{32}}{T (Q_{e^{c}}-Q_{e})} \; , \\
a_{24} &= \frac{N_{e^{c}} a^{e}_{23} (e^{i \phi} \cos\theta + T Q_{e})- N_{e} a^{e}_{32} (e^{i \phi} \cos\theta + T Q_{e^{c}}) }{\sin\theta T (Q_{e^{c}}-Q_{e})} \; , \\
a_{33} &= (a^{u}_{33} N_{u} N_{u^{c}} C_{1} + a^{d}_{33} N_{d} N_{d^{c}} C_{2} + a^{e}_{33} N_{e} N_{e^{c}} C_{3})/(T^2 D_{1})
 \; , \label{eq:a33} \\
a_{34} &= (a^{u}_{33} N_{u} N_{u^{c}} C_{4} + a^{d}_{33} N_{d} N_{d^{c}} C_{5} - a^{e}_{33} N_{e} N_{e^{c}} C_{6})/(T^2 D_{1})  \; , \label{eq:a34} \\
a_{44} &= (a^{u}_{33} N_{u} N_{u^{c}} C_{7} + a^{d}_{33} N_{d} N_{d^{c}} C_{8} - a^{e}_{33} N_{e} N_{e^{c}} C_{9})/(T^2 D_{1})\; , \label{eq:a44}
\end{align}

\noindent with
\vspace{-15pt}
%\begin{eqnarray}
\begin{align}
C_{1} &= Q_{d^c}-Q_{e^c}+Q_d-Q_e ;\\
C_{2} &=  Q_{e^c}-Q_{u^c}+Q_e-Q_u;\\
C_{3} &=  -Q_{d^c}+Q_{u^c}-Q_d+Q_u;\\
C_{4} &=\csc \theta  \left(e^{i \phi } \cos \theta
   \left(Q_{d^c}-Q_{e^c}\right)+Q_d \left(T Q_{d^c}+e^{i \phi
   } \cos \theta \right)-Q_e \left(T Q_{e^c}+e^{i \phi }
   \cos \theta \right)\right);\\
C_{5} &= \csc \theta  \left(Q_e \left(T Q_{e^c}+e^{i \phi } \cos
   \theta \right)+e^{i \phi } \cos \theta
   \left(Q_{e^c}-Q_{u^c}\right)-Q_u \left(T Q_{u^c}+e^{i \phi
   } \cos \theta \right)\right) ;\\
C_{6} &= \csc \theta  \left(Q_d \left(T Q_{d^c}+e^{i \phi } \cos
   \theta \right)+e^{i \phi } \cos \theta
   \left(Q_{d^c}-Q_{u^c}\right)-Q_u \left(T Q_{u^c}+e^{i \phi
   } \cos \theta \right)\right) ;\\
C_{7} &= T^2 Q_d Q_e \csc ^2\theta  Q_{d^c}-T^2 Q_d Q_e \csc
   ^2\theta  Q_{e^c}+T^2 Q_d \csc ^2\theta  Q_{d^c}
   Q_{e^c}-T^2 Q_e \csc ^2\theta  Q_{d^c} Q_{e^c} \nonumber \\ &+e^{2 i
   \phi } \cot ^2\theta  Q_{d^c}+2 T e^{i \phi } Q_d \cot
   \theta  \csc \theta  Q_{d^c}-e^{2 i \phi } \cot
   ^2\theta  Q_{e^c} \nonumber \\ &-2 T e^{i \phi } Q_e \cot \theta  \csc
   \theta  Q_{e^c}+e^{2 i \phi } Q_d \cot ^2\theta -e^{2 i
   \phi } Q_e \cot ^2\theta  ;\\
C_{8} &=e^{2 i \phi } \cot ^2\theta  Q_{e^c}+T^2 Q_e \csc ^2\theta
    Q_u Q_{e^c}-T^2 Q_e \csc ^2\theta  Q_u Q_{u^c}+T^2 Q_e
   \csc ^2\theta  Q_{e^c} Q_{u^c} \nonumber \\ &-T^2 \csc ^2\theta  Q_u
   Q_{e^c} Q_{u^c}+2 T e^{i \phi } Q_e \cot \theta  \csc
   \theta  Q_{e^c}-2 T e^{i \phi } \cot \theta  \csc
   \theta  Q_u Q_{u^c} \nonumber \\ &-e^{2 i \phi } \cot ^2\theta
   Q_{u^c}+e^{2 i \phi } Q_e \cot ^2\theta -e^{2 i \phi }
   \cot ^2\theta  Q_u;\\
C_{9} &= e^{2 i \phi } \cot ^2\theta  Q_{d^c}+T^2 Q_d \csc ^2\theta
    Q_u Q_{d^c}-T^2 Q_d \csc ^2\theta  Q_u Q_{u^c}+T^2 Q_d
   \csc ^2\theta  Q_{d^c} Q_{u^c} \nonumber \\ &-T^2 \csc ^2\theta  Q_u
   Q_{d^c} Q_{u^c}+2 T e^{i \phi } Q_d \cot \theta  \csc
   \theta  Q_{d^c}-2 T e^{i \phi } \cot \theta  \csc
   \theta  Q_u Q_{u^c} \nonumber \\ &-e^{2 i \phi } \cot ^2\theta
   Q_{u^c}+e^{2 i \phi } Q_d \cot ^2\theta -e^{2 i \phi }
   \cot ^2\theta  Q_u ;\\
D_{1} &= Q_d Q_e Q_{d^c}+Q_d Q_{d^c} Q_{e^c}-Q_e Q_{d^c} Q_{e^c}-Q_d
   Q_e Q_{e^c}-Q_d Q_u Q_{d^c}-Q_d Q_{d^c} Q_{u^c} \nonumber \\ &+Q_u Q_{d^c}
   Q_{u^c}+Q_d Q_u Q_{u^c}+Q_e Q_u Q_{e^c}-Q_e Q_u Q_{u^c}+Q_e
   Q_{e^c} Q_{u^c}-Q_u Q_{e^c} Q_{u^c}.
\end{align}
%\end{eqnarray}

%%%%%%%%%%%%%%%%%%%%%%%%%%%%%%%%%%%%%%
%%%%%%%%%%%%%%%%%%%%%%%%%%%%%%%%%%%%%%
\section{Expressions for $a^{\nu}_{33}, a^{R}_{33}$ and $a^{L}_{33}$} \label{App:AppendixB}
In this appendix, we give the expressions for  $a^{\nu}_{33}, a^{R}_{33}, a^{L}_{33}$ for both the $\Phi=45_H$ and $210_H$ cases. Using Eqs.\eqref{eq:af9}, \eqref{eq:a33}, \eqref{eq:a34} and \eqref{eq:a44} it is straightforward to find for the $45_{H}$-case:

\vspace{-10pt}
\begin{equation}\label{anu3345}
a^{\nu}_{33} = a^{u}_{33} \frac{N_{u} N_{u^{c}}}{N_{\nu} N_{\nu^{c}}} + a^{d}_{33} \frac{N_{d} N_{d^{c}}}{N_{\nu} N_{\nu^{c}}} \frac{1+\epsilon /5}{1+\epsilon} - a^{e}_{33} \frac{N_{e} N_{e^{c}}}{N_{\nu} N_{\nu^{c}}} \frac{1+\epsilon /5}{1+\epsilon}.
\end{equation}

\noindent And for the case of $210_{H}$ we find:

\vspace{-10pt}
\begin{align}\label{anu33210}
a^{\nu}_{33} &= a^{u}_{33} \frac{N_u
   N_{u^c}}{N_{\nu } N_{\nu ^c}} \frac{C_{10}}{D_{2}}  +  a^{d}_{33} \frac{N_d N_{d^c}}{N_{\nu } N_{\nu ^c}} \frac{C_{11}}{D_{2}}   + a^{e}_{33}  \frac{N_e N_{e^c}}{ N_{\nu } N_{\nu ^c}} \frac{C_{12}}{D_{2}}  ,
\end{align}

\noindent with \vspace{-12pt}
\begin{align}
C_{10} &= 3 \left(8 \sqrt{6} \epsilon _1^2-4 \left(2 \sqrt{3}
   \epsilon _2+3 \sqrt{2}\right) \epsilon _1+\epsilon _2
   \left(\sqrt{6} \epsilon _2+6\right)\right) , \\
C_{11} &= 3 \left(-8 \sqrt{6} \epsilon _1^2+12 \sqrt{2} \epsilon
   _1+\epsilon _2 \left(\sqrt{6} \epsilon _2-6\right)\right), \\
C_{12} &= 8 \sqrt{6} \epsilon _1^2+4 \left(2 \sqrt{3}
   \epsilon _2+3 \sqrt{2}\right) \epsilon _1-3 \epsilon _2
   \left(\sqrt{6} \epsilon _2+2\right), \\
D_{2} &= \left(4 \epsilon _1-\sqrt{2} \epsilon _2\right)
   \left(2 \sqrt{6} \epsilon _1-3 \sqrt{3} \epsilon _2+3
   \sqrt{2}\right) .
\end{align}

\noindent
Using Eqs. \eqref{eq:a33},  \eqref{eq:a34} and \eqref{eq:a44} for the $45_{H}$-case we have:
\begin{equation}\label{aR3345}
a^{R}_{33} = \frac{3}{2} a^{u}_{33} \frac{N_{u} N_{u^{c}}}{N^{2}_{\nu^{c}}} \frac{1+\epsilon /5}{1+3 \epsilon /5} - \frac{5}{4} a^{e}_{33} \frac{N_{e} N_{e^{c}}}{N^{2}_{\nu^{c}}} \frac{(1+\epsilon /5)^{2}}{\epsilon (1+\epsilon)} +  \frac{5}{4} a^{d}_{33} \frac{N_{d} N_{d^{c}}}{N^{2}_{\nu^{c}}} \frac{1+\frac{3}{5}\epsilon + \frac{3}{25}\epsilon^{2} + \frac{33}{125}\epsilon^{3} }{\epsilon (1+\epsilon) (1+3 \epsilon /5)} .
\end{equation}

\noindent And for $210_{H}$-case we have:
\vspace{-10pt}
\begin{align}\label{aR33210}
a^{R}_{33} &= a^{u}_{33} \frac{N_u N_{u^c}}{N_{\nu ^c}^2} \frac{C_{13} }{D_{3} } -  a^{d}_{33} \frac{N_d N_{d^c}}{N_{\nu ^c}^2} \frac{C_{14} }{D_{3} }  + a^{e}_{33} \frac{N_e N_{e^c}}{N_{\nu ^c}^2} \frac{C_{15} }{D_{3} }  ,
\end{align}

\noindent with
\vspace{-12pt}
\begin{align}
C_{13} &= 18 \sqrt{3} \epsilon _2^4 +36 \sqrt{2} \epsilon _2^3
   -45 \sqrt{6} \epsilon _1 \epsilon _2^3 +12 \sqrt{3} \epsilon _2^2 -156 \epsilon
   _1 \epsilon _2^2   +72 \sqrt{6} \epsilon _1^3
   \epsilon _2 +12 \sqrt{2} \epsilon _1^2 \epsilon
   _2 -30 \sqrt{6} \epsilon _1 \epsilon _2 \nonumber   \\ &+48 \epsilon _1^3 +24 \sqrt{3} \epsilon
   _1^2 , \\
C_{14} &= 9 \sqrt{3} \epsilon _2^4 +18 \sqrt{2} \epsilon _2^3
  -45 \sqrt{6} \epsilon _1 \epsilon _2^3 +72 \sqrt{3} \epsilon _1^2 \epsilon _2^2 +6 \sqrt{3} \epsilon _2^2 -96 \epsilon _1
   \epsilon _2^2   +72 \sqrt{6} \epsilon _1^3 \epsilon
   _2 +36 \sqrt{2} \epsilon _1^2 \epsilon _2 \nonumber \\ &-18 \sqrt{6} \epsilon _1 \epsilon _2 +48
   \epsilon _1^3 +24 \sqrt{3} \epsilon _1^2 , \\
C_{15} &= 16 \sqrt{6} \epsilon _1^3+8 \left(5 \sqrt{3}
   \epsilon _2+6 \sqrt{2}\right) \epsilon _1^2+6 \left(\sqrt{6}
   \epsilon _2^2+8 \epsilon _2+2 \sqrt{6}\right) \epsilon _1-3
   \epsilon _2 \left(3 \sqrt{3} \epsilon _2^2+6 \sqrt{2}
   \epsilon _2+2 \sqrt{3}\right), \\
D_{3} &= 2 \epsilon
   _1 \left(4 \epsilon _1-\sqrt{2} \epsilon _2\right) \left(2
   \sqrt{6} \epsilon _1-3 \sqrt{3} \epsilon _2+3
   \sqrt{2}\right).
\end{align}

\noindent For the case of $45_{H}$ we have:
\vspace{-5pt}
\begin{equation}\label{aL3345}
a^{L}_{33}= \frac{-4}{5} a^{e}_{33} \frac{N_{e^{c}}}{N_{\nu}} \frac{\epsilon}{1+\epsilon} + \frac{1}{2} a^{u}_{33} \frac{N_{u} N_{u^{c}}}{N^{2}_{\nu}} \frac{5+9 \epsilon}{5+3 \epsilon} + \frac{1}{10} a^{d}_{33} \frac{N_{d} N_{d^{c}}}{N^{2}_{\nu}} \frac{25+50 \epsilon + 9 \epsilon^{2}}{(1+\epsilon)(5+3 \epsilon)}.
\end{equation}

\noindent And for the case of $210_{H}$ we have:
\vspace{-5pt}
\begin{align}\label{aL33210}
a^{L}_{33} &= a^{u}_{33} \frac{ N_u N_{u^c}}{ N_{\nu }^2} \frac{C_{16}}{D_{4}} + a^{d}_{33} \frac{ N_d N_{d^c}}{ N_{\nu }^2} \frac{C_{17}}{D_{4}} + a^{e}_{33} \frac{ N_e N_{e^c}}{ N_{\nu }^2} \frac{C_{18}}{D_{4}}  ,
\end{align}

\noindent with
\vspace{-10pt}
\begin{align}
C_{16} &= 3 \left(8 \left(\sqrt{6} \epsilon _2-2\right) \epsilon
   _1^2-4 \left(2 \sqrt{3} \epsilon _2^2+\sqrt{2} \epsilon _2-2
   \sqrt{3}\right) \epsilon _1+\epsilon _2 \left(\sqrt{6}
   \epsilon _2^2+4 \epsilon _2-2 \sqrt{6}\right)\right), \\
C_{17} &= 3 \left(-8 \left(\sqrt{6} \epsilon _2-2\right) \epsilon
   _1^2+\left(12 \sqrt{2} \epsilon _2-8 \sqrt{3}\right)
   \epsilon _1+\epsilon _2 \left(\sqrt{6} \epsilon _2^2-8
   \epsilon _2+2 \sqrt{6}\right)\right), \\
C_{18} &= 8 \sqrt{3} \epsilon _1 \epsilon _2 \left(2 \sqrt{2} \epsilon
   _1-\epsilon _2\right), \\
D_{4} &= 2 \epsilon _2 \left(\sqrt{2} \epsilon _2-4 \epsilon
   _1\right) \left(-2 \sqrt{6} \epsilon _1+3 \sqrt{3} \epsilon
   _2-3 \sqrt{2}\right).
\end{align}

%%%%%%%%%%%%%%%%%%%%%%%%%%%%%%%%%%%%%%
%%%%%%%%%%%%%%%%%%%%%%%%%%%%%%%%%%%%%%
\section{Numerical values of the original Yukawa couplings $a_{ij}$}  \label{App:AppendixC}
In this appendix we present the original coupling matrices $a_{ij}$ for all the different models. $a_{ij}$ matrices are calculated from the fitted parameter sets.

\noindent
Model \textbf{AI}:
\vspace{-5pt}
\begin{equation}
a_{ij} = 10^{-2} \left(
\scalemath{0.9}{
\begin{array}{cccc}
 0.0959072 & 0 & 0.579907\, +0.173698 i & 5.94255\, +2.20933 i \\
 0 & -0.00693205 & 0.134449\, +0.0151987 i & 1.11642\, +0.685207 i \\
 0.579907\, +0.173698 i & 0.134449\, +0.0151987 i & 0.343854\, +2.01413 i &
   16.4068\, +0.693589 i \\
 5.94255\, +2.20933 i & 1.11642\, +0.685207 i & 16.4068\, +0.693589 i &
   192.42\, +53.2691 i \\
\end{array}
}
\right) .
\end{equation}
Model \textbf{AII}:
\vspace{-5pt}
\begin{equation}\label{eq:aij}
a_{ij} = 10^{-2} \left(
\scalemath{0.82}{
\begin{array}{cccc}
 0.127684 & 0 & -0.300661+0.102672 i &
   -2.93065+0.584561 i \\
 0 & -0.00055042 & -0.00799172+0.0297446 i &
   -0.450085+0.185354 i \\
 -0.300661+0.102672 i & -0.00799172+0.0297446 i &
   0.164493\, -2.14188 i & 8.08605\, +16.5832 i \\
 -2.93065+0.584561 i & -0.450085+0.185354 i &
   8.08605\, +16.5832 i & 96.688\, +321.511 i \\
\end{array}
}
\right) .
\end{equation}
Model \textbf{BI}:
\vspace{-1pt}
\begin{equation}
a_{ij} = 10^{-2} \left(
\scalemath{0.87}{
\begin{array}{cccc}
 -0.122115 & 0 & 0.711183\, +0.747142 i & -1.99272-3.01439 i \\
 0 & 0.00569753 & -0.0996008-0.168017 i & 0.22039\, +0.476214 i \\
 0.711183\, +0.747142 i & -0.0996008-0.168017 i & -1.6329-3.89159 i &
   2.08313\, +14.9094 i \\
 -1.99272-3.01439 i & 0.22039\, +0.476214 i & 2.08313\, +14.9094 i &
   10.1819\, -55.9358 i \\
\end{array}
}
\right) .
\end{equation}
Model \textbf{BII}:
\vspace{-1pt}
\begin{equation}
a_{ij} = 10^{-2} \left(
\scalemath{0.87}{
\begin{array}{cccc}
 0.00424453 & 0 & -0.739696+0.100859 i & -0.261416-0.180202
   i \\
 0 & 0.0963929 & 1.59142\, +3.41719 i & -0.546982+1.75381 i
   \\
 -0.739696+0.100859 i & 1.59142\, +3.41719 i & 35.6473\,
   +92.8117 i & -22.3139+45.2696 i \\
 -0.261416-0.180202 i & -0.546982+1.75381 i &
   -22.3139+45.2696 i & -25.416+7.30749 i \\
\end{array}
}
\right) .
\end{equation}
Model \textbf{CI}:
\vspace{-3pt}
\begin{equation}
a_{ij} = 10^{-2} \left(
\scalemath{0.85}{
\begin{array}{cccc}
 0.113952 & 0 & 0.491463\, +0.565896 i & 3.76471\, +4.85316 i \\
 0 & 0.0066433 & -0.0563066+0.107998 i & -0.804669+0.719013 i \\
 0.491463\, +0.565896 i & -0.0563066+0.107998 i & -1.4383+2.14192 i &
   -6.53535+21.208 i \\
 3.76471\, +4.85316 i & -0.804669+0.719013 i & -6.53535+21.208 i &
   -59.2957+144.035 i \\
\end{array}
}
\right) .
\end{equation}
Model \textbf{CII}:
\vspace{-5pt}
\begin{equation}
a_{ij} = 10^{-2} \left(
\scalemath{0.87}{
\begin{array}{cccc}
 -0.104302 & 0 & -0.23862+0.777714 i & -1.41328+3.82589 i \\
 0 & 0.00422339 & -0.0155148+0.115636 i & 0.0287095\, +0.732858 i \\
 -0.23862+0.777714 i & -0.0155148+0.115636 i & -0.0699652+2.31316 i &
   0.347507\, +16.8414 i \\
 -1.41328+3.82589 i & 0.0287095\, +0.732858 i & 0.347507\, +16.8414 i &
   -5.81915+103.489 i \\
\end{array}
}
\right) .
\end{equation}
Model $\rm{\textbf{D}^{a}}$\textbf{I}:
\vspace{-5pt}
\begin{equation}
a_{ij} = 10^{-10} \left(
\scalemath{0.9}{
\begin{array}{cccc}
 -0.00696426 & 0 & -0.245942+0.340282 i &
   -0.981934+1.05034 i \\
 0 & -0.0703536 & -1.2441-0.347061 i & -3.9061-2.2744 i
   \\
 -0.245942+0.340282 i & -1.2441-0.347061 i &
   -11.3476+13.1125 i & -38.737+28.5679 i \\
 -0.981934+1.05034 i & -3.9061-2.2744 i & -38.737+28.5679
   i & -137.986+71.4185 i \\
\end{array}
}
\right).
\end{equation}
Model $\rm{\textbf{D}^{b}}$\textbf{I}:
\vspace{-5pt}
\begin{equation}
a_{ij} = 10^{-2} \left(
\scalemath{0.87}{
\begin{array}{cccc}
 0.0964978 & 0 & 0.0463804 +0.612925 i & 0.23948 -4.0549
   i \\
 0 & -0.00493562 & -0.0692766-0.122098 i & 0.0709726
   +0.757652 i \\
 0.0463804 +0.612925 i & -0.0692766-0.122098 i &
   -1.50677-1.28146 i & 9.26908 +9.93996 i \\
 0.23948 -4.0549 i & 0.0709726 +0.757652 i & 9.26908
   +9.93996 i & -60.1029-54.8455 i
\end{array}
}
\right) .
\end{equation}
Model $\rm{\textbf{D}^{b}}$\textbf{II}:
\vspace{-5pt}
\begin{equation}
a_{ij} = 10^{-2} \left(
\scalemath{0.87}{
\begin{array}{cccc}
 0.00324628 & 0 & 0.00619872 +0.0924098 i & 0.0581533
   +0.495523 i \\
 0 & -0.148375 & 0.234817 +0.306995 i & 2.43678 +3.11641 i
   \\
 0.00619872 +0.0924098 i & 0.234817 +0.306995 i &
   -0.940024-1.19953 i & -10.2704-9.213 i \\
 0.0581533 +0.495523 i & 2.43678 +3.11641 i &
   -10.2704-9.213 i & -63.9596-76.7709 i
\end{array}
}
\right) .
\end{equation}
Model \textbf{EI}:
\vspace{-5pt}
\begin{equation}
a_{ij} = 10^{-2} \left(
\scalemath{0.89}{
\begin{array}{cccc}
 0.00565532 & 0 & -0.266188+0.0701303 i & 0.518805\, -0.533918 i \\
 0 & 0.10865 & 0.425618\, +1.82085 i & -3.18677-2.87341 i \\
 -0.266188+0.0701303 i & 0.425618\, +1.82085 i & -14.7905+7.46579 i &
   14.094\, -32.9412 i \\
 0.518805\, -0.533918 i & -3.18677-2.87341 i & 14.094\, -32.9412 i &
   19.4372\, +73.7643 i \\
\end{array}
}
\right).
\end{equation}
Model \textbf{EII}:
\vspace{-5pt}
\begin{equation}
a_{ij} = 10^{-2} \left(
\scalemath{0.87}{
\begin{array}{cccc}
 -0.00491523 & 0 & -0.113267+0.379235 i & -0.0710291-0.772947 i \\
 0 & -0.0989956 & 1.91708\, +0.433922 i & -4.21969+0.992353 i \\
 -0.113267+0.379235 i & 1.91708\, +0.433922 i & -14.9942+2.32607 i &
   31.5765\, -18.8751 i \\
 -0.0710291-0.772947 i & -4.21969+0.992353 i & 31.5765\, -18.8751 i &
   -50.6035+73.4838 i \\
\end{array}
}
\right) .
\end{equation}
Model \textbf{FI}:
\vspace{-3pt}
\begin{equation}
a_{ij} = 10^{-2} \left(
\scalemath{0.89}{
\begin{array}{cccc}
 -0.101322 & 0 & 1.00796\, -1.0692 i & -3.40853+1.66912 i \\
 0 & 0.00628798 & -0.117644+0.20702 i & 0.5911\, -0.454689 i \\
 1.00796\, -1.0692 i & -0.117644+0.20702 i & -3.59013+11.2045 i & 20.5757\,
   -20.2831 i \\
 -3.40853+1.66912 i & 0.5911\, -0.454689 i & 20.5757\, -20.2831 i &
   -63.553+23.8071 i \\
\end{array}
}
\right).
\end{equation}
Model \textbf{FII}:
\vspace{-3pt}
\begin{equation}
a_{ij} = 10^{-2} \left(
\scalemath{0.9}{
\begin{array}{cccc}
 0.00631171 & 0 & -0.0913116-0.224948 i & 0.708328\, +0.52168 i \\
 0 & -0.106855 & 0.477564\, +1.57686 i & -4.01776-3.01492 i \\
 -0.0913116-0.224948 i & 0.477564\, +1.57686 i & 12.5721\, -5.02818 i &
   -18.1583+34.5582 i \\
 0.708328\, +0.52168 i & -4.01776-3.01492 i & -18.1583+34.5582 i &
   -22.102-106.709 i \\
\end{array}
}
\right).
\end{equation}

\end{appendices}

%%%%%%%%%%%%%%%%%%%%%%%%%%%%%%%%%%%%%%%%%%%%%
%%%%%%%%%%%%%%%%%%%%%%%%%%%%%%%%%%%%%%%%%%%%%
%\newpage
\FloatBarrier

\end{document}